\documentclass[iop,revtex4,twocolumn,apj,numberedappendix,appendixfloats]{emulateapj}
\usepackage{apjfonts}
\usepackage{amsmath} 
\usepackage{booktabs} 

\shorttitle{Galaxy Pairs in MaNGA} 
\shortauthors{Fu et al.}
\journalinfo{The Astrophysical Journal}
\submitted{accepted by ApJ on 2018 February 27}

\usepackage[pagebackref=false,letterpaper=true,colorlinks=true,citecolor=blue,linkcolor=blue,breaklinks=true,bookmarks=true]{hyperref}

\newcommand{\kms}{km~s$^{-1}$}
\newcommand{\msun}{$M_{\odot}$}
\newcommand{\msunyr}{$M_{\odot}~{\rm yr}^{-1}$}
\newcommand{\lsun}{$L_{\odot}$}
\newcommand{\ergs}{erg~s$^{-1}$}

\newcommand{\nod}{\nodata}

\begin{document}

\title{SDSS-IV MaNGA: Galaxy Pair Fraction and Correlated Active Galactic Nuclei}

\author{
Hai Fu\altaffilmark{1}, 
Joshua L. Steffen\altaffilmark{1}, 
Arran C. Gross\altaffilmark{1}, 
Y. Sophia Dai\altaffilmark{2}, 
Jacob W. Isbell\altaffilmark{1}, 
Lihwai Lin\altaffilmark{3}, 
David Wake\altaffilmark{4}, 
Rui Xue\altaffilmark{1}, 
Dmitry Bizyaev\altaffilmark{5,6},
Kaike Pan\altaffilmark{5}
}
\altaffiltext{1}{Department of Physics \& Astronomy, The University of Iowa, 203 Van Allen Hall, Iowa City, IA 52242, USA}
\altaffiltext{2}{Chinese Academy of Sciences South America Center for Astronomy (CASSACA), National Astronomical Observatories, Chinese Academy of Sciences, Beijing 100012, China}
\altaffiltext{3}{Institute of Astronomy \& Astrophysics, Academia Sinica, Taipei 10617, Taiwan}
\altaffiltext{4}{School of Physical Sciences, The Open University, Milton Keynes, MK7 6AA, UK}
\altaffiltext{5}{Apache Point Observatory and New Mexico State University, P.O. Box 59, Sunspot, NM, 88349-0059, USA}
\altaffiltext{6}{Sternberg Astronomical Institute, Moscow State University, Moscow, Russia}

\begin{abstract}
We have identified 105 galaxy pairs at $z \sim 0.04$ with the MaNGA integral-field spectroscopic data. The pairs have projected separations between 1~kpc and 30~kpc, and are selected to have radial velocity offsets less than 600~\kms\ and stellar mass ratio between 0.1 and 1. The pair fraction increases with both the physical size of the integral-field unit and the stellar mass, consistent with theoretical expectations. We provide the best-fit analytical function of the pair fraction and find that $\sim$3\% of $M^*$ galaxies are in close pairs. For both isolated galaxies and paired galaxies, active galactic nuclei (AGN) are selected using emission-line ratios and H$\alpha$ equivalent widths measured inside apertures at a fixed physical size. We find AGNs in $\sim$24\% of the paired galaxies and binary AGNs in $\sim$13\% of the pairs. To account for the selection biases in both the pair sample and the MaNGA sample, we compare the AGN comoving volume densities with those expected from the mass- and redshift-dependent AGN fractions. We find a strong ($\sim$5$\times$) excess of binary AGNs over random pairing and a mild ($\sim$20\%) deficit of single AGNs. The binary AGN excess increases from $\sim$2$\times$ to $\sim$6$\times$ as the projected separation decreases from $10-30$~kpc to $1-10$~kpc. Our results indicate that pairing of galaxies preserves the AGN duty cycle in individual galaxies but increases the population of binary AGNs through correlated activities. We suggest tidally-induced galactic-scale shocks and AGN cross-ionization as two plausible channels to produce low-luminosity narrow-line-selected binary AGNs.
\end{abstract}


\section{Introduction} \label{sec:intro}

In the dark energy plus cold dark matter ($\Lambda$CDM) universe, massive galaxies are built upon a series of major and minor mergers. Cosmological hydrodynamical simulations estimate that about 10\% of the stellar mass in a Milky-Way-sized galaxy today \citep[$M_{\rm halo} \simeq 10^{12}$~\msun;][]{McMillan11} is accreted from other galaxies (i.e., formed {\it ex-situ}) and the fraction of ex-situ stellar mass increases to $\sim$60\% for ten times more massive halos \citep{Pillepich17a}. Furthermore, gas-rich major mergers offer an attractive physical origin to the correlation between the mass of the supermassive black hole (SMBH) and the velocity dispersion of the stellar bulge \citep[e.g.,][]{Kormendy13} and the similarity in the cosmic histories of star formation and black hole accretion \citep[e.g.,][]{Shankar09}. Clearly, galaxy mergers are a key element in galaxy evolution. Here we focus on two important parameters that can be obtained by observations: the galaxy pair fraction and the level of merger-induced SMBH accretion.

How frequently galaxies merge at different cosmic epochs is a fundamental measurement to constrain cosmological simulations. One of the methods to estimate the merger rate is to measure the fraction of close galaxy pairs and convert it to the merger rate by dividing the merging timescales from $N$-body simulations \citep[e.g.,][]{Boylan-Kolchin08}. Previous studies have selected merging pairs primarily using spectroscopic redshifts \citep[e.g.,][]{Patton02,Lin04,Xu04,De-Propris05,Lin08,Bundy09,Keenan14,Robotham14}. Because of spectroscopic incompleteness in close pairs (as we will discuss below), photometric redshifts have also been used to increase the sample size, especially at higher redshifts \citep[e.g.,][]{Kartaltepe07,Xu12}. At low redshift ($z \sim 0.1$), the consensus is that only $\sim$2\% of $M^*$ galaxies \citep[$M^* = 4.6\times10^{10}$~\msun;][]{Baldry12} are involved in major mergers with projected separations less than 30~kpc and the corresponding merger rate is $\sim$0.04 per $M^*$ galaxy per Gyr \citep{Robotham14}. But how strongly the merger rate evolves with redshift remains under debate \citep[e.g.,][]{Man16}.

Merger-driven nuclear activities have long been suspected, because simulations show that gravitational torques from tidally-induced stellar bars can drive large amounts of interstellar gas to the central kpc in a short timescale \citep{Barnes96}. However, the galactic-scale bars are insufficient to feed the SMBH to form an active galactic nucleus (AGN) because the gas still needs to be driven from kpc to sub-pc scales to reach the accretion disk. A promising mechanism is ``bars-within-bars'' --- strong gravitational torques from a series of gravitational instabilities at different scales continuously remove the angular momentum of the gas until it reaches the black hole's accretion disk \citep{Shlosman89}. But it requires running a series of simulations on successively smaller scales to explore these instabilities \citep{Hopkins10}. As a result, state-of-the-art merger simulations and cosmological simulations of galaxy evolution still compute black hole growth using ``sub-grid'' models, such as the modified \citet{Bondi44} formula for a spherical accretion flow \citep[e.g.,][]{Volonteri15,Weinberger17a}. Therefore, observations of on-going mergers are needed to better constrain the key small-scale physical processes in galaxy mergers. Indeed, the AGN fraction of close galaxy pairs exceeds that of field galaxies when the projected separations decrease below $\sim$50~kpc, as shown by a number of studies based on a range of AGN selection methods: e.g., optical emission lines \citep[e.g.,][]{Ellison11,Liu12}, X-ray thermal continuum \citep[e.g.,][]{Silverman11,Koss12}, mid-IR colors \citep[e.g.,][]{Satyapal14}, and radio synchrotron emission \citep[e.g.,][]{Fu15b}. These studies suggest that mergers are able to trigger and even synchronize SMBH accretion. 

However, traditional spectroscopic surveys have three limitations that affect the selection of close galaxy pairs and the study of AGN activities in mergers: (1) constant magnitude limits for spectroscopic targets, (2) fiber/slit collisions, and (3) fixed angular apertures. For instance, the Sloan Digital Sky Survey \citep[SDSS;][]{York00} only targets objects brighter than $r < 17.8$ and avoids observing objects simultaneously if their angular separation is less than 55\arcsec\ due to fiber collision \citep{Blanton03a}. Therefore, a galaxy pair with projected separations less than 55\arcsec\ will have complete spectroscopic information only if (1) the system is in overlapping regions of adjacent tiles (which accounts $\sim$30\% of the SDSS footprint), and (2) both components are brighter than $r < 17.8$. Furthermore, the fixed angular diameter of the optical fibers \citep[2\arcsec\ or 3\arcsec;][]{Smee13} samples varying physical sizes over a range of redshift, causing artificial trends in all aperture-related quantities such as specific star formation rates (sSFR) and emission line ratios. These problems can be resolved by surveys using large-format integral-field units (IFUs), which offer spatially resolved spectroscopy covering an area that is significantly larger than the target galaxy. In this paper, we use integral-field spectroscopic data from the MaNGA survey \citep[Mapping Nearby Galaxies at Apache Point Observatory;][]{Bundy15} to set a local benchmark on the galaxy pair fraction and revisit the question of how mergers affect AGN activity. The two previous studies of galaxy pairs, selected using integral-field spectroscopy, have a total sky coverage of 12.4~arcmin$^2$ and have focused on mergers of star-forming galaxies at high redshifts \citep{Lopez-Sanjuan13,Ventou17}. In comparison, the MaNGA IFUs have covered a total area of $\sim$300~arcmin$^2$ and have observed both star-forming and quiescent galaxies between $0.01 < z < 0.15$. Here we focus on galaxy pairs for which both members are fully covered by a single MaNGA IFU.

The paper is structured as follows. In Section \ref{sec:manga}, we describe the MaNGA data products, properties of the sample, and our spectral fitting method. In Section \ref{sec:pair}, we identify close galaxy pairs in the MaNGA sample and model the selection biases to determine the intrinsic pair fraction at $z \sim 0.04$. In Section \ref{sec:agn}, we perform emission-line classification and measure the AGN fraction as a function of stellar mass and redshift for a control sample of isolated galaxies. In Section \ref{sec:agn_in_pairs}, we investigate whether and how galaxy mergers affect AGN activity through a detailed comparison between paired galaxies and the control sample. In Section \ref{sec:summary}, we conclude with a summary and a discussion on scenarios that can explain the observed excess of correlated AGN activity in close pairs. Throughout we adopt the AB magnitude system and a flat $\Lambda$CDM cosmology with $\Omega_{\rm m}=0.27$, $\Omega_\Lambda=0.73$ and $H_0$ = 70 km~s$^{-1}$~Mpc$^{-1}$ (so that $h = H_0/100~{\rm km~s}^{-1}~{\rm Mpc}^{-1} = 0.7$).

\section{MaNGA Data and Analysis} \label{sec:manga}

\subsection{Galaxy Sample} \label{sec:weights}

MaNGA utilizes 17 optical fiber-bundle IFUs to observe thousands of nearby galaxies with the SDSS 2.5-meter telescope \citep{Drory15}. For galaxy observations, there are five IFU sizes optimized to the target galaxy size distribution: the 19, 37, 61, 91, and 127-fiber bundles fill hexagonal areas with long-axis diameters of 12.5\arcsec, 17.5\arcsec, 22.5\arcsec, 27.5\arcsec, and 32.5\arcsec, respectively. A three-point dithering pattern is employed to improve the spatial uniformity and it increases the field-of-view by $\sim$3.5\arcsec\ for all IFUs \citep{Law15}. The raw data are reduced by a dedicated Data Reduction Pipeline \citep[DRP;][]{Law16}, which produces fully calibrated, stacked datacubes for each IFU, with a pixel size of 0.5\arcsec$\times$0.5\arcsec$\times$10$^{-4}$~dex in ($\alpha$, $\delta$, $\log \lambda$). 

As described in \citet{Wake17}, the MaNGA main galaxy sample is selected from an enhanced version of the NASA-Sloan Atlas catalog (NSA v1\_0\_1; \url{http://www.nsatlas.org}) which includes 641,490 galaxies at $z < 0.15$. The target catalog contains 41,154 galaxies over 7,362~deg$^2$ with a redshift range between $0.01 < z < 0.15$ and a luminosity range of $-17.7 < \mathcal{M} < -24.0$, where $\mathcal{M}$ is the rest-frame $i$-band absolute magnitude inside an elliptical Petrosian aperture. It is a combination of three subsamples --- the Primary ($\sim$47\%), the Color Enhanced ($\sim$16\%), and the Secondary ($\sim$37\%). The Primary and the Color Enhanced subsamples together are called the ``Primary$+$'' sample, for which $\sim$78\% of the allocated IFUs would cover out to 1.5 times the effective radii of target galaxies ($R_e$), which is the semi-major axis of the ellipse that contains half of the elliptical Petrosian flux. For the Secondary sample, $\sim$75\% of the allocated IFUs would cover out to 2.5$R_e$. The Fourteenth Public Data Release \citep[DR14;][]{Abolfathi17} includes 2,772 galaxy datacubes, 54 of which are repeated observations. The MaNGA sample in this paper consists of the 2,618 unique galaxies in the main galaxy sample. By selecting only the main galaxy sample, we have excluded the 100 galaxies that are either from ancillary programs or from the commissioning runs.

The MaNGA sample is designed to achieve a relatively flat stellar mass distribution between $9.5 < \log(M/M_\odot) < 11.5$ and a semi-uniform spatial coverage out to 1.5$R_e$ (Primary$+$ sample) or 2.5$R_e$ (Secondary sample). As a result, more luminous galaxies are selected at higher redshifts (to fit within the IFUs) and over wider redshift ranges (to compensate the declining luminosity function) than less luminous galaxies. The sample is thus distributed on two narrow stripes in the parameter space of absolute magnitude ($\mathcal{M}$) and redshift ($z$). For simplicity, we would refer to this distribution as the ``$\mathcal{M}-z$ correlation''. Therefore, the comoving volume (hereafter ``volume'' in short) of the survey is a function of the absolute magnitude. To recover a volume-limited sample, we need to use the 1/$V_{\rm max}$ weights \citep{Schmidt68}. For example, the volume density of MaNGA galaxies between $\mathcal{M}_{\rm min}$ and $\mathcal{M}_{\rm max}$ can be calculated in the following way:
\begin{equation}
\begin{aligned}
n(\mathcal{M}_{\rm min},\mathcal{M}_{\rm max}) &\equiv \int_{\mathcal{M}_{\rm min}}^{\mathcal{M}_{\rm max}} \Phi(\mathcal{M}) d \mathcal{M} \\
	&\equiv \int_{\mathcal{M}_{\rm min}}^{\mathcal{M}_{\rm max}} \frac{d N/d \mathcal{M}}{V_{\rm tiled}[z_{\rm min}(\mathcal{M}),z_{\rm max}(\mathcal{M})]} d\mathcal{M} \\
	&= \sum_{j = 1}^{N_{\rm tiled}} \frac{1}{V_{\rm tiled}[z_{\rm min}(\mathcal{M}_{j}),z_{\rm max}(\mathcal{M}_{j})]}
\end{aligned}
\end{equation}
where $\Phi(\mathcal{M})$ is $i$-band luminosity function, $N_{\rm tiled}$ is the total number of galaxies satisfying the MaNGA sample selection within the 7,362~deg$^2$ survey area when tiled with 1,800 plates, and $V_{\rm tiled}[z_{\rm min}(\mathcal{M}),z_{\rm max}(\mathcal{M})]$ is the volume covered by the full survey area and the redshift range corresponding to the absolute magnitude $\mathcal{M}$. Because only a subset of galaxies will be observed ($N_{\rm obs} < N_{\rm tiled}$), in practice we should always use the incompleteness-corrected volumes:
\begin{equation}
n(\mathcal{M}_{\rm min},\mathcal{M}_{\rm max}) \simeq \sum_{j = 1}^{N_{\rm obs}} \frac{N_{\rm tiled}/N_{\rm obs}}{V_{\rm tiled}[z_{\rm min}(\mathcal{M}_{j}),z_{\rm max}(\mathcal{M}_{j})]} 
\end{equation}
where the ratio on the numerator is used to scale the volume of the full survey to the effective area of the completed part of the survey. Following \citet{Wake17}, we define the dimensionless $1/V_{\rm max}$ weights for each galaxy as:
\begin{equation} \label{eq:weight}
W_j \equiv \frac{N_{\rm tiled}}{N_{\rm obs}} \frac{10^6~{\rm Mpc}^3}{V_{\rm tiled}[z_{\rm min}(\mathcal{M}_{j}),z_{\rm max}(\mathcal{M}_{j})]}
\end{equation}
so that,
\begin{equation} \label{eq:sumweight}
n(\mathcal{M}_{\rm min},\mathcal{M}_{\rm max}) \simeq \sum_{j = 1}^{N_{\rm obs}} W_j/10^6~{\rm Mpc}^3
\end{equation}

As part of the DR14, the MaNGA target catalog provides $W_j$ for each of the subsamples and four sample combinations \citep[see the Appendix of][]{Wake17}. These weights are calculated using a hypothetical observed sample resulted from adaptively tiling 1,800 plates. However, MaNGA will complete only $\sim$600 plates in its survey lifetime, and DR14 completed only $\sim$9\% of the simulated sample with 163 unique plates. Hence, we have re-calculated these weights for our adopted cosmology and the actually observed sample in DR14. Note that using the weights is especially important when the galaxies in a subset cover a significant range in absolute magnitudes or redshift.

\subsection{Spectral Fitting} \label{sec:spfit}

We have developed an IDL package SPFIT to fit the MaNGA spectra produced by the DRP.
The package is built upon the Penalized Pixel-Fitting method (pPXF) of \citet{Cappellari04} and simultaneously fits emission lines and stellar continuum with a nonlinear optimizer. Following the standard procedure, we model the observed spectrum as a superposition of emission lines and simple stellar populations (SSPs), after correcting for the Galactic extinction and de-redshifting to the rest frame (i.e., the redshift needs to be known {\it a priori}). We use the SSP library of MIUSCAT \citep{Vazdekis12}, which have a wide spectral range (3465$-$9469~\AA) with a uniform spectral resolution (FWHM = 2.5~\AA). The SSPs are matched to the MaNGA spectral resolution and are convolved with the line-of-sight velocity distribution (LOSVD). Intrinsic reddening of the stellar continuum is modeled with either a \citet{Calzetti00} extinction law or a high-order multiplicative polynomial. Both the LOSVD and the profile of the emission lines are parameterized as separate Gauss-Hermite series \citep{van-der-Marel93} to the fourth order. Following \citet{Cappellari04}, we penalize all of the $h_3$ and $h_4$ terms of the Gauss-Hermite series with an additional bias term in the residual calculation to stabilize the fit.  We tie the kinematics of all emission lines to minimize the number of free parameters. Favoring its fast speed, we use the Levenberg-Marquard nonlinear least-squares minimization algorithm \citep[][\S15.5]{Press92} implemented in MPFIT \citep{Markwardt09}. But for complex models like ours, the success of the fitting routine relies on a good set of initial guesses. The pPXF method is robust because it solves the weights of the templates with a linear algorithm \citep[][]{Lawson74} {\it independently} from solving the Gauss-Hermite LOSVD with a nonlinear optimizer (MPFIT). We thus adopt a three-stage fitting procedure: First, we mask out spectral regions around possible emission lines from the input spectrum and use pPXF on the masked spectrum with SSP-only templates to obtain the initial ``best-fit'' parameters of the stellar continuum; Secondly, we subtract the best-fit stellar continuum model from the spectrum and use pPXF to fit the residual emission-line-only spectrum with Gaussian emission-line templates to obtain the initial ``best-fit'' parameters of the emission lines; Finally, we use MPFIT on the input spectrum with the two sets of ``best-fit'' parameters from pPXF to simultaneously fit all of the parameters describing the emission lines and the stellar continuum. The final best-fit model from this simultaneous fit improves the $\chi^2$ from the two-step pPXF model and allows the emission line gas and stellar populations to have separate kinematics. 

We estimate the 1$\sigma$ uncertainties of the emission line fluxes based on their best-fit amplitude-to-noise ratios (A/N) and the observed line widths ($\sigma$ in unit of pixels). Although MPFIT computes the statistical errors of the free parameters by evaluating the covariance matrix, it tends to underestimate the errors because the kinematics of the emission lines have been tied together. Re-fitting individual emission lines without any tied parameters would be computationally expensive and would not provide robust results at low S/N. On the other hand, the fractional error of the line flux ($\delta f/f = (S/N)^{-1}$) is mostly determined by the A/N and the width of the line. With a suite of synthetic Gaussian emission lines, we estimate $\delta f/f$ as a function of A/N and $\sigma$ in the range between $2 < A/N < 20$ and 1~pix~$< \sigma < 20$~pix. For each emission line in the data, we then use this pre-computed table and the observed A/N and line width to evaluate its flux uncertainty.  

\section{Close Galaxy Pairs in MaNGA} \label{sec:pair}

\subsection{Pair Selection} \label{sec:pair_selection}

In this section, we select spectroscopically confirmed (i.e., kinematic) galaxy pairs in the MaNGA sample. We are interested in pairs for which both members are covered by a {\it single} IFU, so that we have complete spectroscopic information. The sample does not include galaxy pairs partially covered by the MaNGA IFUs. A few percent of the sample are triples or higher multiples, but we do not attempt to discuss them separately because of the small sample size and their complex selection biases. The pair selection procedure has two stages: (1) pair candidate selection based on the SDSS images and the photometric catalog and (2) spectral confirmation of the pair candidates using the MaNGA datacubes. 

\begin{enumerate}

\item We begin the first stage by downloading the SDSS cutout images and the DR9 photometric catalog \citep{Ahn12} for the 2,618 MaNGA fields. For each field, we keep all of the galaxy-type photometric objects at least 1\arcsec\ inside the {\it dithered} hexagonal IFU boundary.   

\item Here we define the galaxy mass ratio as $\mu$ = $M_2/M_1$, where $M_2$ is always less than $M_1$ so that $0 < \mu < 1$. Typically, major mergers and minor mergers are defined as $\mu \geq 1/3$ and $0.1 \leq \mu < 1/3$, respectively. As an attempt to select major and minor mergers, we identify the MaNGA fields that contain objects with $r$-band magnitude differences less than $\Delta r < 2.5$ relative to the MaNGA target. The $r$-band flux density ratio is used to approximate the mass ratio, assuming a constant mass-to-light ratio. However, the catalog magnitudes may not be reliable for close galaxy pairs \citep{Simard11}, because of the source segmentation issues of the SDSS photometric pipeline when dealing with crowded fields. As a work-around, we take a conservative approach by choosing the smallest $\Delta r$ calculated with all three available magnitudes in the catalog (Model, PSF, and Petrosian magnitudes). It is expected that pairs with $\mu < 0.1$ could have been selected because of problematic photometry. To remove them, in the next stage, we use stellar masses estimated from MaNGA spectra to re-calculate the mass ratios.

\item We overlay the photometric catalog on the SDSS cutout images for the 439 galaxy pair candidates to remove isolated single galaxies that are over-deblended by the photometric pipeline. A total of 236 candidates survived the visual inspection. 

\item The remaining 2,179 MaNGA fields do not show companions with $\Delta r < 2.5$ according to the photometric catalog. To search for close pairs that were not deblended by the photometric pipeline, three of us (HF, AG, and JS) inspected the cutout images overlaid with the catalog sources to look for significant sources that were not in the catalog. We identified 8 additional pairs and include them in the pair candidate sample. 

\end{enumerate}

The pair candidate sample selected above contains a total of 244 fields. Note that we have used the SDSS photometric catalog instead of the NSA catalog to select pair candidates because only 13 of the 2,618 MaNGA fields contain NSA objects other than the MaNGA targets. All of the 13 pairs from the NSA are also selected with the SDSS photometric catalog. The NSA catalog misses most of the pair candidates because ({\it a}) it only includes galaxies with previously known spectroscopic redshifts, which most of the secondaries in our pairs do not have due to fiber collision (see \S~\ref{sec:collision}), and ({\it b}) the NSA photometry algorithm is optimized for large nearby galaxies, so it requires a much higher threshold to deblend a child from a parent object when compared to the photometric pipeline \citep[i.e., the under-deblending issue;][]{Blanton11}.

Physically unrelated pairs due to line-of-sight projection are common and must be removed. The subsequent spectral confirmation stage starts from the pair candidates and uses spectroscopic information offered by the MaNGA datacubes to remove projected pairs.

\begin{enumerate}

\item We begin by extracting spectra for both components in each pair candidate. Here we adopt a 2\arcsec-diameter circular aperture to match with the fiber diameter and the typical seeing. We fit the spectra assuming both components in a MaNGA field are at the same target redshift. If the input redshift is roughly correct, the best-fit models from SPFIT would provide the stellar mass, stellar and gas kinematics, and emission line fluxes. 

\item We then remove projected pairs by inspecting individual spectrum to check the quality of the spectral fit. As expected, incorrect input redshift results in poor model fits and spectral features indicating different redshifts are usually evident. A total of 95 pairs are rejected in this process, including 35 galaxy-star pairs, 45 projected galaxy pairs with large redshift differences ($\Delta z \gg 0.01$), and 15 ambiguous cases where the SNR of the fainter component is too low to detect any spectral features. 

\end{enumerate}

There remain 149 spectroscopically confirmed galaxy pairs. Finally, we refine the pair sample by incrementally excluding the following sources: 

\begin{itemize}

\item 30 pairs with stellar mass ratios of $\mu < 0.1$. Here we use the stellar masses from the best-fit models of the nuclear spectra to weed out pairs that passed the cut in $r$-band flux density ratio because of problematic photometry.

\item 6 pairs with $\Delta v > 600$~\kms. Higher radial velocity offsets may indicate greater distances along the line-of-sight. Although the 600~\kms\ threshold is a subjective choice, our main results are unchanged if we adopt a lower threshold at 300~\kms. Of course, there are larger statistical uncertainties due to the 16\% smaller sample. 

\item 8 pairs with $\Delta \theta > \theta_{\rm IFU}$. We define $\theta_{\rm IFU}$ as the short-axis radius of the IFU, which is the radius of the inscribed circle that is 1\arcsec\ away from the dithered boundary. The 19, 37, 61, 91, and 127-fiber IFUs have $\theta_{\rm IFU}$ = 5.93\arcsec, 8.09\arcsec, 10.26\arcsec, 12.42\arcsec, and 14.59\arcsec, respectively. For $\sim$99\% of the MaNGA observations, the target is placed at the center of the IFU. This requirement avoids selection bias depending on the position angle of the pair, because these excluded pairs can only be selected when the companions are located near one of the six corners of a hexagon. One of the excluded pairs, 8553-9102, is a binary AGN where the secondary AGN is at the lower-left corner, 13.2\arcsec\ or 18.6~kpc away from the primary. Additionally, this requirement excludes wide-separation pairs that are in our sample only because the primary target is offset from the IFU center.

\end{itemize}

The final sample includes 105 close pairs with mass ratios between $0.1 < \mu < 1$, radial velocity separations of $\Delta v < 600$~\kms, angular separations between 2.4\arcsec~$< \Delta \theta < $~14.3\arcsec, and projected separations between 1.2~kpc~$< r_p <$~29.4~kpc. There are 56 ($\sim$53\%) major pairs ($\mu > 1/3$) and 49 ($\sim$47\%) minor pairs ($0.1 < \mu < 1/3$). The sample does not include any pairs with $\Delta \theta \lesssim 2\arcsec$ because of limitations due to atmosphere seeing. For each galaxy pair, we define the MaNGA target as the {\it primary} component and its companion as the {\it secondary} component, and the term ``pair'' refers to both components. For most fields, the primary is placed at the center of the IFU.

\subsection{Potential Issues from Fiber Collision} \label{sec:collision}

There are two types of fiber collisions that may affect the pair sample --- the fiber collision from the SDSS I/II spectroscopy \citep{Blanton03a} and the IFU collision from MaNGA itself \citep{Law15}. The collision radii of SDSS I/II spectroscopy and MaNGA IFUs are 55\arcsec\ and 120\arcsec, respectively. The collisions prevent simultaneous observations of multiple galaxies within a circle of the collision radius. Because the second component in a conflicted pair can only be observed if it happens to be in a tile overlapping region, the fiber collision decreases the spectroscopic completeness of paired galaxies, and the IFU collision decreases the IFU allocation
rate of such galaxies. 

The overall spectroscopic completeness of the NSA catalog is 98.6\% for galaxies with $13 < r < 17$, thanks to the inclusion of redshifts from a number of spectroscopic surveys. But the NSA is still affected by the fiber collision issue of the SDSS: the spectroscopic completeness decreases to $\sim$69\% for similarly bright galaxies in apparent pairs with $\Delta\theta < 55\arcsec$ \citep{Wake17}. This is consistent with the $\sim$30\% tile overlapping fraction in SDSS \citep{Blanton03a}: suppose $\sim$98.6\% of the first component and $\sim$30\% of the second component are observed, one would expect a spectroscopic completeness $\sim$64\% for all paired galaxies. In addition, because the brightest galaxy in a collision area is more likely to be selected for spectroscopic observations in the first pass, the primary is more massive than the secondary in most (96/105) of our pairs.

However, in contrast to pairs selected using single-fiber spectroscopy, neither of the two collisions affects our pair sample based on the MaNGA IFU data, because our selection requires only one component in a pair to have a spectroscopic redshift and that the pair must be covered in a single IFU ($\Delta\theta < \theta_{\rm IFU}$). It is thus unimportant whether both components in a pair have spectroscopic redshifts and whether they both get allocated IFUs in MaNGA. Although the spectroscopic completeness is only $\sim$69\% for galaxies in pairs with $\Delta\theta < 55\arcsec$, for essentially all of these pairs at least one component has acquired a spectroscopic redshift, thus meeting our first selection criterion. The {\it effective} spectroscopic completeness of galaxy pairs is thus the same as the overall spectroscopic completeness (98.6\%). The IFU collision does not affect the pair sample either because we require a pair to be fully covered by a single IFU.

\subsection{Galaxy Pair Fraction} \label{sec:fpair}

\begin{figure}[!tb]
\epsscale{1.2}
\plotone{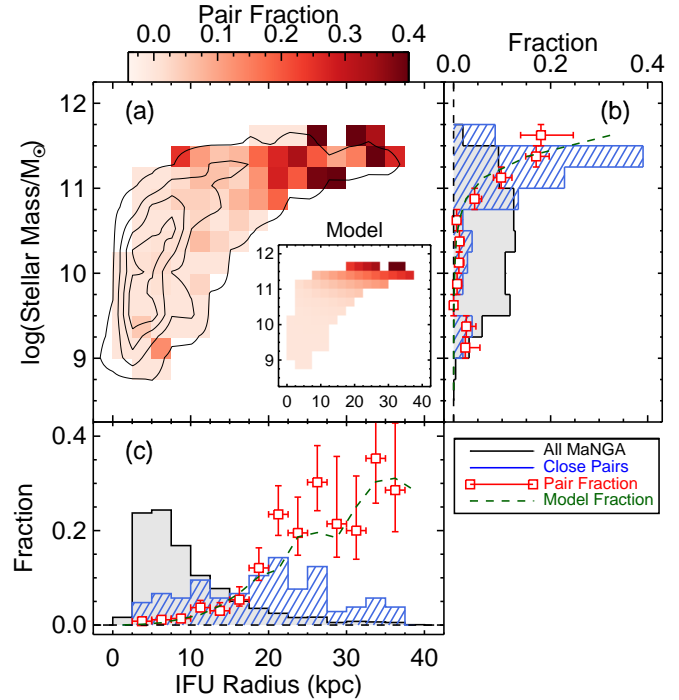}
\caption{Selection biases of the pair sample. Panel ($a$) shows the observed pair fraction as a function of the IFU radius at the target redshift ($R_{\rm IFU}$) and stellar mass ($M$). For galaxy pairs, $M$ refers to the total mass of the system. The contours show the distribution of the MaNGA sample at $dN_{\rm gal} = (6,34,62,90)$ per cell. The inset shows the best-fit model to the observed pair fractions (Eq.~\ref{eq:fpairmod}), restricted to the same region in the parameter space as occupied by MaNGA galaxies. In panels ($b$) and ($c$), we show the normalized one-dimensional distributions of the pair sample ({\it blue hatched histogram}) compared to those of the MaNGA main galaxy sample ({\it gray filled histogram}), the observed pair fractions in each bin ({\it red data points with error bars}), and the pair fractions computed from the model ({\it green dashed curves}). 
\label{fig:fpair}} 
\end{figure}

Taking at face value, $\sim$4\% (105/2618) of MaNGA galaxies are in close pairs. But this does {\it not} directly translate to a pair fraction for low-redshift galaxies because the MaNGA sample is not volume-limited. In addition, the pair sample have important selection biases because (1) the variable IFU size at the target redshift determines the maximum projected pair separation, and (2) the pair fraction should depend on stellar mass and redshift. Quantifying the selection biases is thus needed to recover the intrinsic pair fraction.

First of all, the finite IFU size introduces an important selection bias because our pair selection requires both components in a pair to be covered by the same IFU. We define the radius of the IFU at the target redshift $z$ as $R_{\rm IFU} = \theta_{\rm IFU} \times D_A(z)$, where $\theta_{\rm IFU}$ is the IFU radius defined in \S~\ref{sec:pair_selection} and $D_A(z)$ is the angular diameter distance at redshift $z$. The IFU radius ranges between 1.6~kpc~$< R_{\rm IFU} <$~38~kpc over the redshift range of the sample ($0.01 < z < 0.15$). Obviously, the companions of MaNGA targets, if they exist, can only be found in IFUs whose radii are larger than the projected separations (i.e., $\theta_{\rm IFU} \geq \Delta \theta$). For a given stellar mass for the primary galaxy and a mass ratio $\mu$, the pair fraction should increase linearly with $R_{\rm IFU}$, which defines the maximum projected separation. This is because the dynamical friction time is linearly proportional to the initial separation when there is mass loss due to tidal stripping, $\tau_{\rm fric} \propto r_0$, a result derived using Chandrasekhar's dynamical friction formula \citep[][\S8.1.1]{Binney08} and verified in $N$-body simulations \citep[e.g.,][]{Jiang14}. 
Secondly, the MaNGA sample is designed to have a flat stellar mass distribution between $9.5 < \log(M/M_\odot) < 11.5$. The pair fraction should increase significantly over this stellar mass range because more massive galaxies are more strongly clustered \citep[e.g.,][]{Zehavi02,Cooray02}. 
Lastly, the MaNGA sample spans a redshift range between $0.01 < z < 0.15$ with a median redshift of $\bar{z} = 0.04$. But the pair fraction should only increase mildly, by less than $\sim$30\%, over this redshift range, given that $f_{\rm pair} \propto (1+z)^\alpha$ with $\alpha$ between 1.5 and 2.0 \citep{Hopkins10b}. Redshift is thus a less important parameter when compared to $R_{\rm IFU}$ and $M$.

Therefore, we compute the pair fraction in the plane of $R_{\rm IFU}$ and $\log M$. The stellar masses in the NSA catalog are computed from the elliptical Petrosian magnitudes using the $K$-correct package \citep{Blanton07}. The two-dimensional (2D) pair fraction is defined as 
\begin{equation}
f_{\rm pair}(R_{\rm IFU},M) = dN_{\rm pair}/dN_{\rm gal}
\end{equation}
where $dN_{\rm pair}$ and $dN_{\rm gal}$ are the numbers of close pairs and MaNGA targets in each cell of $dR_{\rm IFU}\times d\log M =  2.5~{\rm kpc}\times0.25~{\rm dex}$, respectively. The 1$\sigma$ uncertainty of the pair fraction is then estimated from the Poisson errors of $dN_{\rm pair}$ and $dN_{\rm gal}$. Because the galaxies in each cell have essentially the same stellar mass, the number ratio here is equivalent to the ratio of the $1/V_{\rm max}$ weights defined in \S~\ref{sec:weights}. 

The 2D pair fraction is shown in the main panel of Fig.~\ref{fig:fpair}. The MaNGA sample is distributed in a banana-shaped area in the parameter space, as shown by the contours. The distribution reflects the $\mathcal{M}-z$ correlation of the MaNGA sample. In the side panels, we compare the 1D distributions of the MaNGA sample and our close pair sample, as well as the 1D pair fractions. As expected, the pair sample is biased to large IFU sizes and high stellar masses, and a model assuming linear dependences on both parameters offers a good fit to the data. The best-fit single-parameter model below gives a reduced $\chi^2$ of 0.7 and is shown in the inset of Fig.~\ref{fig:fpair}$a$:  
\begin{equation}
f_{\rm pair}^{\rm mod}(R_{\rm IFU},M) = 9.3\pm1.3\% \Big(\frac{R_{\rm IFU}}{30~{\rm kpc}}\Big) \Big(\frac{M}{10^{11}~M_\odot}\Big) \label{eq:fpairmod}
\end{equation}
where the 1$\sigma$ uncertainty corresponds to the interval of $\chi^2 - \chi^2_{\rm min} = 1.0$.
We note that the 1D distributions of pairs along each of the two parameters are determined by {\it not only} the physically motivated bias function above {\it but also} the distribution of the parent sample in the parameter space. Using the best-fit model (Eq.~\ref{eq:fpairmod}) and the 2D distribution of MaNGA galaxies, we can reproduce the observed 1D pair fractions ({\it dashed green curves} in Fig.~\ref{fig:fpair}$b$ \& $c$). 

Alternatively, we can fit the observed 2D pair fraction using the halo-occupation-based (HOB) model of merger rates \citep{Hopkins10b}. We use the semi-empirical model to compute the expected pair fraction in each $dR_{\rm IFU}\times d\log M$ cell for mergers with $\mu > 0.1$ and all gas fractions. We set the default stellar mass functions, $R_{\rm IFU}$ as the maximum projected separation, and a fixed redshift at $\bar{z} = 0.04$. We obtain essentially the same result when replacing the fixed redshift with the mean redshift in each cell, consistent with the expectation that the merger fraction evolves slowly with redshift at $z < 0.15$. The HOB model gives the following relation between pair fraction and $R_{\rm IFU}$ and $M$:
\begin{equation}
f_{\rm pair}^{\rm HOB}(R_{\rm IFU},M) \propto R_{\rm IFU} M^{\beta(M)} \label{eq:fpairHOB}
\end{equation}
where the power-law slope $\beta(M)$ increases from $\sim$0.2 to $\sim$1.7 as the stellar mass increases from $10^{9}$~\msun\ to $10^{11.5}$~\msun. The average slope of $\bar{\beta} \sim 1.0$ is consistent with the assumption in the previous model (Eq.~\ref{eq:fpairmod}). But the HOB model systematically underestimates the pair fraction. To provide an adequate fit to the data, we could {\it either} multiply 2-3$\times$ to the predicted pair fractions {\it or} add 0.2-0.3~dex to the mean stellar mass in each bin. The former is comparable to the quoted factor of $\sim$2 uncertainty of the model \citep{Hopkins10b}, and the latter is consistent with the typical 0.3~dex systematic uncertainty in stellar mass estimates \citep{Blanton07}, showing that the level of disagreement is within the expectation.

We can use the best-fit model in Eq.~\ref{eq:fpairmod} and the observed stellar mass function to estimate the {\it cumulative} pair fraction ($F_{\rm pair}$) in a volume-limited sample for any given ranges of projected separation and stellar mass\footnote{Note that the model is only calibrated within $r_p < 40$~kpc and $9.0 < \log M/M_\odot < 11.5$.}. Alternatively, we can use the $1/V_{\rm max}$ weights (\S~\ref{sec:weights}) to estimate $F_{\rm pair}$ in MaNGA: 
\begin{equation}
F_{\rm pair} = \frac{\sum_{j = 1}^{N_{\rm gal}} W_j f_{\rm pair}^{\rm mod}(r_{\rm max}, M_j)}{\sum_{j = 1}^{N_{\rm gal}} W_j} 
\end{equation}
where $r_{\rm max}$ is the maximum projected separation for the pair selection. For $r_{\rm max} = 20 h^{-1}$~kpc = 28.6~kpc and $M^*$ galaxies ($10.16 < \log M/M_\odot < 12.16$, i.e., $\log M^* \pm 0.5$~dex), we obtain $F_{\rm pair} = 3.4\pm0.5\%$. Given that $\sim$53\% of our pairs are major pairs with $\mu > 1/3$, our result agrees well with previous estimates of the major-merger pair fraction for $M^*$ galaxies at $z < 0.2$ \citep[$F_{\rm major} \sim 1-4\%$ for $r_{\rm max} = 20h^{-1}$~kpc; see the compilation in][]{Robotham14}.

\begin{figure*}[!tb]
\epsscale{1.2}
\plotone{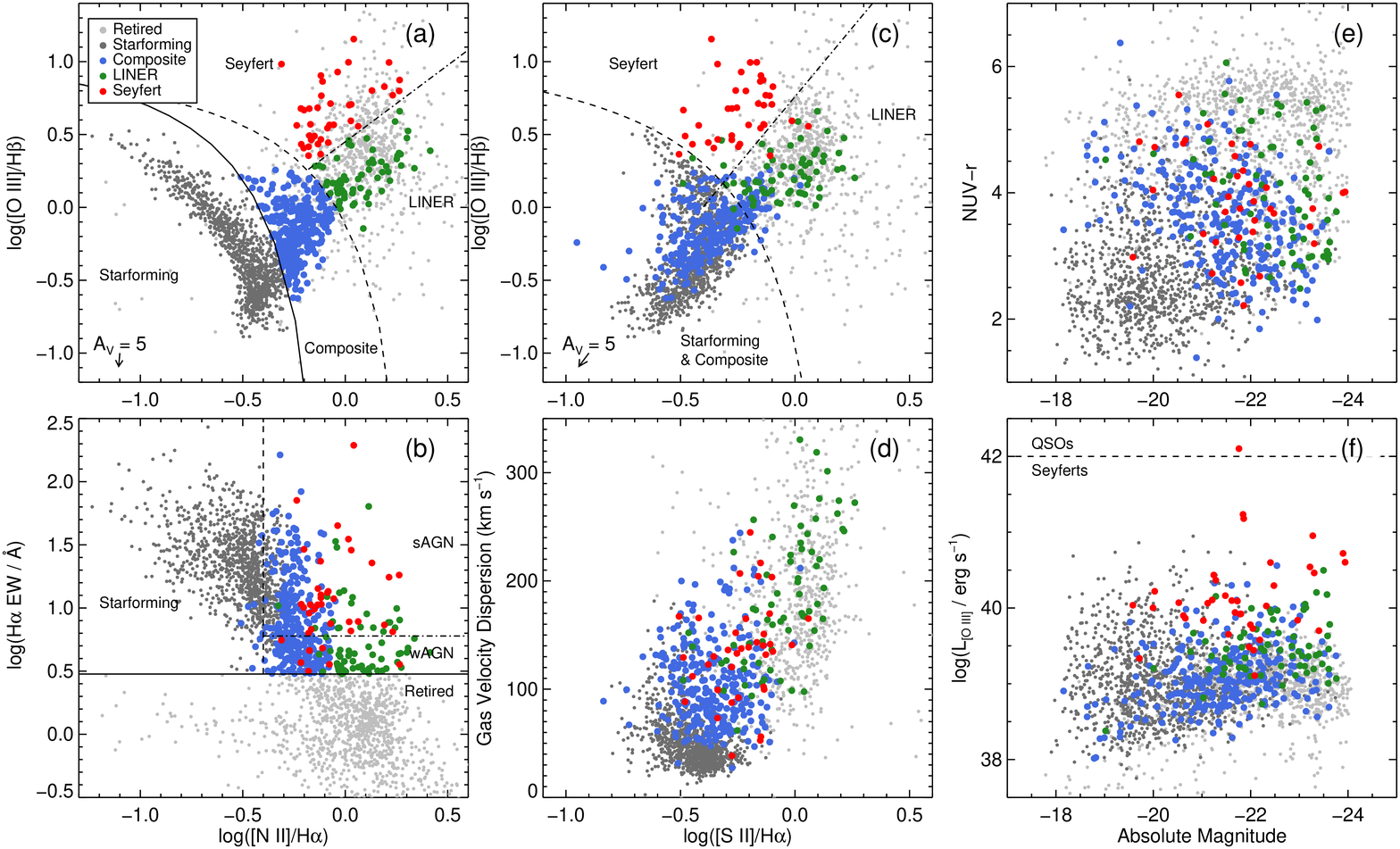}
\caption{
Distributions of different emission-line galaxies in MaNGA. All emission-line measurements are from nuclear spectra extracted with 2.6~kpc-diameter apertures. The symbols are color-coded according to the emission-line classification described in \S~\ref{sec:class}. The retired galaxies, star-forming galaxies, starburst-AGN composites, LINERs, and Seyferts are plotted in {\it light gray, dark gray, blue, green}, and {\it red}, respectively. 
({\it a}) the [N\,{\sc ii}] BPT diagram -- log([O\,{\sc iii}]$\lambda$5007/H$\beta$) vs. log([N\,{\sc ii}]$\lambda$6584/H$\alpha$).
({\it b}) the WHAN diagram -- H$\alpha$ equivalent width vs. log([N\,{\sc ii}]$\lambda$6584/H$\alpha$).
({\it c}) the [S\,{\sc ii}] BPT diagram -- log([O\,{\sc iii}]$\lambda$5007/H$\beta$) vs. log([S\,{\sc ii}]$\lambda\lambda$6716,6731/H$\alpha$).
({\it d}) emission-line velocity dispersion (km/s) vs. log([S\,{\sc ii}]$\lambda\lambda$6716,6731/H$\alpha$).
({\it e}) the color magnitude diagram -- GALEX/NUV$-$SDSS/$r$ color vs. the $i$-band absolute magnitude ($\mathcal{M}$). 
({\it f}) [O\,{\sc iii}]\,$\lambda$5007 line luminosity vs. $\mathcal{M}$. The horizontal dashed line separates Seyferts and QSOs.
All magnitudes have been $K$-corrected to the rest frame.
The reddening arrows in ($a$) and ($c$) point from the observed to the intrinsic values for $A_{V} = 5$~mag, assuming the reddening curve of \citet{Cardelli89} with $R_V = 3.1$.
\label{fig:bptwhan}} 
\end{figure*}

\section{AGNs in MaNGA} \label{sec:agn}

\subsection{Emission-Line Classification} \label{sec:class}

For emission-line classification, we extract a nuclear spectrum from each MaNGA datacube with a 2.6~kpc-diameter circular aperture. The extraction apertures are defined by the centroid coordinates and the redshift of the MaNGA target galaxy from the NSA catalog. The spectra within the aperture are coadded and fit with SPFIT (\S~\ref{sec:manga}). We define the extraction aperture using a fixed physical scale instead of a fixed angular scale (as in \S~\ref{sec:pair}), because over the redshift range between $0.01 < z < 0.15$ the angular scale increases from 0.22~kpc/arcsec to 2.62~kpc/arcsec. And we choose a diameter of 2.6~kpc because (1) it corresponds to half of the fiber diameter (1\arcsec) at the highest redshift and (2) the smallest IFUs with $\theta_{\rm IFU} = 5.93\arcsec$ are just large enough to cover the aperture at the lowest redshift.

We classify the nuclear spectra using a combination of the [O\,{\sc iii}]$\lambda$5007/H$\beta$ vs. [N\,{\sc ii}]$\lambda$6584/H$\alpha$ BPT diagram \citep{Baldwin81} and the H$\alpha$ Equivalent Width (EW) vs. [N\,{\sc ii}]/H$\alpha$ WHAN diagram \citep{Cid-Fernandes11}. 

In the [N\,{\sc ii}]/H$\alpha$ BPT diagram in Fig.~\ref{fig:bptwhan}{\it a}, we overlay the \citet{Kewley01} theoretical maximum starburst line (Eq.~\ref{eq:Kew01}; {\it dashed curve}) and the \citet{Kauffmann03} empirical separation line between star-forming galaxies and AGNs (Eq.~\ref{eq:Kau03}; {\it solid curve}):
\begin{equation}\label{eq:Kew01}
\log(\rm{[O\,\textsc{iii}]}/\rm{H}\beta) = \frac{0.61}{\log(\rm{[N\,\textsc{ii}]}/\rm{H}\alpha)-0.47}+1.19
\end{equation}
\begin{equation}\label{eq:Kau03}
\log(\rm{[O\,\textsc{iii}]}/\rm{H}\beta) = \frac{0.61}{\log(\rm{[N\,\textsc{ii}]}/\rm{H}\alpha)-0.05}+1.30
\end{equation}
Galaxies above the maximum starburst line are either Seyferts or Low-Ionization Nuclear Emission-Line Regions (LINERs), below the empirical star-forming/AGN dividing line are star-forming galaxies, and galaxies in between are starburst-AGN composite galaxies. LINERs and Seyferts can be approximately separated by the empirical separation line below \citep[][{\it dash-dotted line} in Fig.~\ref{fig:bptwhan}$a$]{Schawinski07}: 
\begin{equation}\label{eq:Sch07}
\log (\rm{[O\,\textsc{iii}]} / \rm{H}\beta) = 1.05 \log(\rm{[N\,\textsc{ii}]}/\rm{H}\alpha) + 0.45
\end{equation}

The ``AGN branch'' in the BPT diagram consists of starburst-AGN composites, LINERs, and Seyferts, where presumably AGN photoionzation have important contributions to the enhanced forbidden line luminosities. However, the hard ionization spectrum of evolved stellar populations can also produce line ratios along the AGN branch \citep[e.g.,][]{Stasinska08}. \citet{Cid-Fernandes11} showed that the H$\alpha$ EW is a robust indicator of the dominance of post-AGB stars, because it strongly correlates with the ratio between the observed and the expected H$\alpha$ luminosity of photoionization from stellar populations older than 100~Myr (when post-AGBs start to dominate the UV radiation field). They suggest an empirical threshold of EW(H$\alpha$) = 3~\AA\ to separate ``retired galaxies (RGs)''  where evolved stellar population is the dominant ionization mechanism from other emission-line galaxies where the emission lines trace gas ionized by star formation, AGN, or radiative shocks. Hence, we classify all galaxies with EW(H$\alpha$)~$<$~3\AA\ as RGs regardless of the emission line ratios, and we use dividing lines in the [N\,{\sc ii}]/H$\alpha$ BPT diagram to classify the emission-line galaxies with EW(H$\alpha$)~$\geq$~3\AA. Note that the H$\alpha$ EW cut removed essentially all of the galaxies with insufficient S/N in any of the four strong emission lines in the BPT diagram.

The above classification scheme cannot be applied to two rare groups of objects. So we have to treat them separately:
\begin{enumerate}

\item The [O\,{\sc iii}]$\lambda$5007 line at $z \simeq 0.114$ falls on top of the strong telluric line at 5577~\AA, making it impossible to measure its flux. We thus classify the emission-line galaxies with EW(H$\alpha$)~$\geq$~3\AA\ between $0.113 < z < 0.116$ using the WHAN diagram, where objects with EW(H$\alpha$)~$\geq$~3\AA\ can be subdivided into star-forming galaxies, weak AGNs (wAGN), and strong AGNs (sAGN) based on the [N\,{\sc ii}]/H$\alpha$ ratio and H$\alpha$ EWs (Fig.~\ref{fig:bptwhan} ({\it b})). There are two such cases (8244-12704 and 8454-6102) and both are classified as sAGN and we put them in the LINER category.

\item The emission-line measurements are unreliable when there are broad emission lines. So we identify type-1 Seyferts by inspecting the nuclear spectra and cross-matching the MaNGA sample with the SDSS DR7 type-1 AGN catalog of \citet{Stern12a}.  

\end{enumerate}

Out of the 2,618 MaNGA galaxies, we find 1,171 RGs, 1,051 SFGs, 271 starburst-AGN composites, 68 LINERs, 38 type-2 Seyferts, and 14 type-1 Seyferts. Five galaxies are unclassified because of (1) incorrect redshifts in the NSA catalog (8461-6101, 8551-9101), (2) foreground stars near the galactic nuclei (8155-12702, 8623-12703), and (3) an incorrect IFU position (8239-3701). We exclude these five unclassified galaxies in the following discussion.

We define a broad AGN sample that include type-1 AGNs and all galaxies with EW(H$\alpha$)~$\geq$~3\AA\ and above the \citet{Kauffmann03} dividing line. The sample of 391 AGNs is comprised of 69.3\% starburst-AGN composites, 17.4\% LINERs, 9.7\% type-2 Seyferts, and 3.6\% type-1 Seyferts. In the last four panels of Figure~\ref{fig:bptwhan}, we show the distribution of the three narrow-line AGN subclasses in the parameter spaces covered by [S\,{\sc ii}]/H$\alpha$ line ratio, emission-line velocity dispersion, rest-frame $NUV$-$r$ color, $i$-band absolute magnitude, and [O\,{\sc iii}] luminosity. 

The [O\,{\sc iii}]$\lambda$5007 luminosity is an indicator of the AGN bolometric luminosity. Fig.~\ref{fig:bptwhan}$f$ shows that most of the MaNGA AGNs are low-luminosity AGNs. The [O\,{\sc iii}] luminosities range between $10^{38} \lesssim L_{\rm [O~III]} \lesssim 10^{42}$~\ergs, or AGN bolometric luminosities between $10^{8} \lesssim L_{\rm bol} \lesssim 10^{12}$~\lsun, given the bolometric correction factor of 3,500 \citep{Heckman04}. All except one MaNGA AGN are below the QSO-Seyfert dividing line at $L_{\rm bol} = 10^{12}$~\lsun\ \citep{Brusa07}, which corresponds to a black hole accretion rate of 0.6~\msunyr\ for a radiative efficiency of $\epsilon = 0.1$. Note that the [O\,{\sc iii}] luminosities measured inside apertures of a constant physical size do {\it not} correlate with the absolute magnitudes of the host galaxies, in contrast to line luminosities measured inside apertures of a fixed angular size, indicating that the apparent correlation is introduced by the redshift-dependent aperture correction factor. 

On the other hand, it is evident from the color-magnitude diagram (Fig.~\ref{fig:bptwhan}$e$) that AGN host galaxies are biased to high stellar mass and intermediate colors between the red sequence and the blue cloud, confirming previous results using single-fiber SDSS spectra \citep{Kauffmann03}. The stellar-mass bias naturally explains the high emission-line velocity dispersions of AGNs (Fig.~\ref{fig:bptwhan}$d$), which are comparable to the stellar velocity dispersions measured from absorption features. The high velocity dispersion does not necessarily imply the importance of shocks. In the next subsection, we will quantify the bias of the AGN host galaxies.

\subsection{Host Galaxies of MaNGA AGNs} \label{sec:fagn}

\begin{figure}[!tb]
\epsscale{1.2}
\plotone{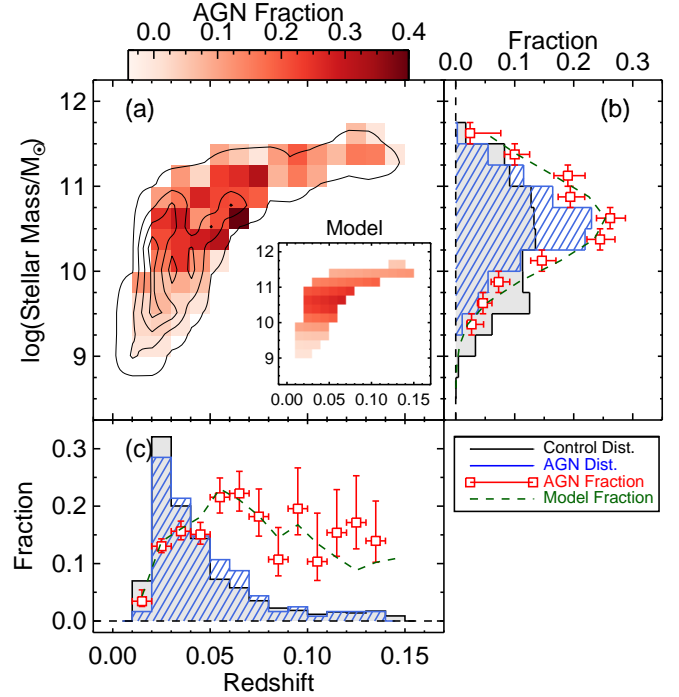}
\caption{Similar to Fig.~\ref{fig:fpair} but for the AGN fraction in the control sample as a function of redshift and stellar mass. The model in the inset is given in Eq.~\ref{eq:fagn}. The contours show the distribution of the control sample at $dN_{\rm ctrl} = (8,46,84,122)$ per cell.
\label{fig:fagn}} 
\end{figure} 

\begin{figure*}[!t]
\epsscale{1.2}
\plotone{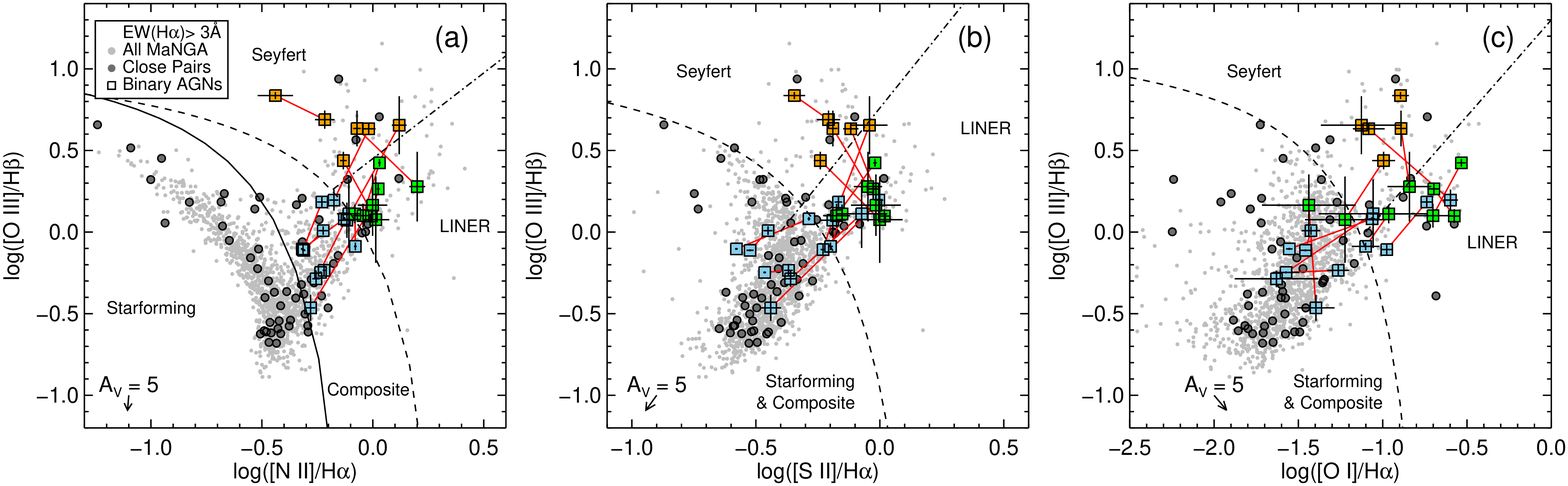}
\caption{The BPT diagnostic diagrams for paired galaxies. 
({\it a}) log([O\,{\sc iii}]$\lambda$5007/H$\beta$) vs. log([N\,{\sc ii}]$\lambda$6584/H$\alpha$).
({\it b}) log([O\,{\sc iii}]/H$\beta$) vs. log([S\,{\sc ii}]$\lambda\lambda$6716,6731/H$\alpha$).
({\it c}) log([O\,{\sc iii}]/H$\beta$) vs. log([O\,{\sc i}]$\lambda$6300/H$\alpha$).
Large data points with 1$\sigma$ error bars are members of the 14 binary AGNs. They are color-coded by the emission-line classification based on the [N\,{\sc ii}]/H$\alpha$ diagram --- starburst-AGN composites, LINERs, and Seyferts are blue, green, and orange, respectively. Red solid lines connect the two components belonging to the same pair. For comparison, large dark gray circles and small light gray circles show the other emission-line galaxies in the pair sample and the control sample, respectively. Retired galaxies are not plotted. 
\label{fig:bptbagn}} 
\end{figure*}

As an attempt to separate between stochastic and interaction-induced activities, we define a control sample of 2,508 ``isolated'' galaxies by excluding the 105 close pairs (and the five unclassified galaxies) from the MaNGA sample. The control sample inevitably includes a small number of interacting galaxies partially covered by the MaNGA IFUs. Given the small fraction of pairs at low redshift, this contamination should be insignificant. In fact, our results remains the same even if we include the close pairs in the control sample. 

Previous studies of AGN host galaxies have shown that the AGN fraction is primarily a function of stellar mass and redshift. It does not depend on the size of the IFU as the pair fraction does. In Fig.~\ref{fig:fagn} we show the AGN fraction in the control sample as a function of the two parameters. We define the 2D AGN fraction as: 
\begin{equation}
f_{\rm AGN}(M, z) = dN_{\rm AGN}/dN_{\rm ctrl}
\end{equation}
where $dN_{\rm AGN}$ and $dN_{\rm ctrl}$ are the numbers of AGNs and MaNGA targets in each cell of $dz\times d\log M =  0.01\times0.25~{\rm dex}$, respectively; both are counted within the control sample. Similar to the 2D pair fraction in Fig.~\ref{fig:fpair}, the distribution of the parent sample is determined by the $\mathcal{M}-z$ correlation intrinsic to the MaNGA sample design. The figure shows that the AGN fraction strongly peaks at intermediately high stellar mass ($\sim4\times10^{10}$~\msun). The dependence on redshift is less apparent because of the limited redshift range of the sample. Assuming a log-normal function in stellar mass and a power-law evolution in redshift, the following model provides a good fit to the observed AGN fraction:
\begin{equation}
f_{\rm AGN}^{\rm mod}(M,z) = f_0 \exp\Big[-\Big(\frac{\log M/M_\odot - b}{2\sigma}\Big)^2\Big] (1+z)^4 \label{eq:fagn}
\end{equation}
where $f_0 = 0.22\pm0.02$, $b = 10.57\pm0.05$, and $\sigma = 0.54\pm0.05$. Because our data do not provide a strong constraint on the redshift evolution, we have fixed the power-law index to 4 to be consistent with the steep evolution of the AGN luminosity function at $z \lesssim 1$ \citep[][]{Ueda03}. Since the model is not limited to the stellar mass range covered by the MaNGA target sample at a given redshift, it can be used to estimate the probability that a secondary galaxy in a pair is an AGN. We will use this model to estimate the expected volume densities of AGNs in the pair sample and compare them with the observed volume densities in the next section. 

\section{AGNs in Galaxy Pairs} \label{sec:agn_in_pairs}

\subsection{Emission-Line Classification} \label{sec:class_pairs}

Our sample of close galaxy pairs includes 105 pairs (\S~\ref{sec:pair}). To be consistent with the control sample, we re-extract the nuclear spectra from these 210 paired galaxies with 2.6~kpc-diameter apertures. Six of the 105 pairs have projected separations less than 2.6~kpc, so we set the aperture diameters to their projected separations to avoid aperture overlapping. 

We apply the same emission-line classification scheme as described in \S~\ref{sec:class} and find 125 (59.5\%) RGs, 35 (16.7\%) SFGs, and 50 (23.8\%) AGNs. The AGNs include 28 starburst-AGN composites, 13 LINERs, 8 type-2 Seyferts, and 1 type-1 Seyfert. Table~\ref{tab:pairs} lists the properties of the pairs and Fig.~\ref{fig:bptbagn} shows the emission-line ratios of the paired galaxies in the three commonly used BPT diagrams. The distribution of paired galaxies in different emission-line classes is significantly different from that of the control sample. Out of the 2,508 galaxies in the control sample, there are 1,108 (44.1\%) RGs, 1,035 (41.2\%) SFGs, and 365 (14.5\%) AGNs (including 257 starburst-AGN composites, 60 LINERs, 34 type-2 Seyferts, and 14 type-1 Seyferts). The increased fraction of RGs and the decreased fraction of SFGs in paired galaxies are both likely a result of the stellar mass biases of the pair sample (see Fig.~\ref{fig:fpair}). In the next subsection, we will show that the high AGN fraction in paired galaxies is also primarily due to the sample bias.

We identify 14 binary AGNs where both the primary and the secondary are classified as AGN. They account for $\sim$13\% of the pair sample. The first section of Table~\ref{tab:pairs} lists the properties of the binary AGNs and Fig.~\ref{fig:bpt2d} shows their SDSS images and spatially resolved emission-line measurements from the MaNGA datacubes. The projected separations of the binary AGNs range between $4~{\rm kpc} < r_p < 21~{\rm kpc}$, and 10 out of the 14 have $r_p < 10~{\rm kpc}$. The components of the binary AGNs are highlighted in Fig.~\ref{fig:bptbagn} with color symbols. Six binaries have different AGN subclasses between primary and secondary (two cases in each of the three combinations), while eight show consistent AGN subclasses (five starburst-AGN-Composite pairs, two LINER pairs, and 1 Seyfert pair). Four of the 14 binary AGNs were previously identified. The pairs in 8083-9101, 8133-12704, and 8612-12705 have SDSS DR7 single-fiber spectra for both components, so they have been previously identified as AGN pairs by \citet{Liu11}. Note that, although 8549-12705 and 9049-12701 are also in the AGN pair sample of \citet{Liu11}, they are not classified as binary AGNs here because of their low H$\alpha$ EWs (i.e., they are RGs). Additionally, 8329-6102 was identified by mid-IR colors and was confirmed with {\it Chandra} X-ray observations \citep{Ellison17}.

The rest of the pair sample includes 16 AGN-RG pairs, 6 AGN-SFG pairs, 7 RG-SFG pairs, 51 RG-RG pairs, and 11 SFG-SFG pairs. The last two groups account for 59\% of the pair sample. Besides random pairing, some fraction of those could be galaxies with correlated star-formation activity or inactivity (RGs). Given the overall SFG fractions of 16.7\% in the pair sample, we expect only three random SFG-SFG pairs in a sample of 105 pairs. Similarly, we would expect $\sim$37 random RG-RG pairs. This simple calculation suggests that there could be a significant population of correlated star-forming galaxies and retired galaxies. But a more rigorous investigation requires quantifying the stellar-mass dependence of star-forming and retired fractions in the control sample (similar to Fig.~\ref{fig:fagn} and Eq.~\ref{eq:fagn} for AGNs) and utilizing the 1/$V_{\rm max}$ weights described in \S~\ref{sec:weights} to compare the observed and expected volume densities. In the next subsection, we will carry out such an investigation on correlated AGNs. The same method can be applied to correlated star-forming galaxies and retired galaxies.

\subsection{Enhanced and Correlated AGN Activities} \label{sec:agn_excess}

Here we investigate (1) whether paired galaxies have a higher AGN duty cycle than isolated galaxies and (2) whether there is a significant fraction of correlated AGNs in galaxy pairs. The former can be addressed by measuring the AGN fraction in paired galaxies and compare it with that of the control sample. The latter can be addressed using the incidence of binary AGNs. However, the two separate processes are closely related and both can be attributed to tidal interactions. Because correlated AGNs could also increase AGN fraction in paired galaxies, we should avoid correlated AGNs when estimating the AGN fraction. On the other hand, elevated AGN fraction in paired galaxies would also lead to more binary AGNs, so we need to consider both random pairing and correlated AGNs when modeling the binary AGN fraction.  

At first glance, AGNs do appear to be more common in the pair sample than in the control sample: 22\% (23/105) of the primaries and 26\% (27/105) of the secondaries are AGNs, while only 14.5\% (365/2508) of the control sample are AGNs. The high AGN fractions may suggest enhanced SMBH accretion in interacting pairs. However, it could also be due to the selection bias of the pair sample. Recall that the pair sample is biased to high stellar mass and large IFU size (consequently, high redshift) when compared to the parent sample (Fig.~\ref{fig:fpair}) and the AGN fraction varies with stellar mass and redshift (Fig.~\ref{fig:fagn} and Eq.~\ref{eq:fagn}). On the other hand, the number of binary AGNs (14/105 or 13\%) is significantly higher than expected from random pairing: one would expect only $\sim$6 binaries ($=105\times22\%\times26\%$) even given the high AGN fractions in pairs. It thus suggests that the majority of the binary AGNs may be produced by correlated AGN activity.

First, we test whether sample biases can explain the high AGN fractions in paired galaxies by comparing the observed and the expected volume densities of uncorrelated AGNs. Both volume densities are calculated using the $1/V_{\rm max}$ weights, which effectively convert MaNGA into a volume-limited survey (see \S~\ref{sec:weights}). To avoid correlated AGNs, we select two samples of AGNs in paired galaxies: (1) 36 pairs where either the primary or the secondary is an AGN (hereafter ``either-AGNs'') and (2) 22 pairs where only one member is an AGN (hereafter ``single AGNs''). Note that the either-AGN sample includes binary AGNs but avoids correlated AGNs because each binary AGN is counted only once, which is in contrast to a sample that counts all of the AGNs in paired galaxies. The single-AGN sample excludes binary AGNs completely but has a smaller sample size. Using Eq.~\ref{eq:sumweight}, the observed volume densities (in units of $10^{-6}~{\rm Mpc}^{-3}$) of each sample are respectively:
\begin{equation} \label{eq:n1agn_obs}
n_{\rm eagn}^{\rm obs} = \sum_{j=1}^{N_{\rm eagn}} W_j ~~~\&~~~ n_{\rm sagn}^{\rm obs} = \sum_{j=1}^{N_{\rm sagn}} W_j 
\end{equation}
where $N_{\rm eagn}$ and $N_{\rm sagn}$ are the number of either-AGNs and single AGNs in each bin, respectively, and $W_j$ is the $1/V_{\rm max}$ weight of the MaNGA target as defined in Eq.~\ref{eq:weight}. If we interpret the AGN fraction of the control sample, $f_{\rm agn}(M,z)$, as the probability that any galaxy with stellar mass $M$ at redshift $z$ would host an AGN, we can estimate the expected volume densities of the above two samples of uncorrelated AGNs:
\begin{equation} \label{eq:n1agn_exp}
\begin{aligned}
n_{\rm eagn}^{\rm exp} &= \sum_{j=1}^{N_{\rm pair}} W_j~f_{\rm agn}(M_{j}^p,z_j)+\sum_{j=1}^{N_{\rm pair}} W_j~f_{\rm agn}(M_{j}^s,z_j) \\ &- \sum_{j=1}^{N_{\rm pair}} W_j~f_{\rm agn}(M_{j}^p,z_j)~f_{\rm agn}(M_{j}^s,z_j)\\
n_{\rm sagn}^{\rm exp} &= \sum_{j=1}^{N_{\rm pair}} W_j~f_{\rm agn}(M_{j}^p,z_j)~[1-f_{\rm agn}(M_{j}^s,z_j)] \\ &+ \sum_{j=1}^{N_{\rm pair}} W_j~f_{\rm agn}(M_{j}^s,z_j)~[1-f_{\rm agn}(M_{j}^p,z_j)] \\
\end{aligned}
\end{equation}
where $N_{\rm pair}$ is the number of galaxy pairs in each bin, $M_{j}^p$ and $M_{j}^s$ are respectively the stellar mass of the primary and that of the secondary, and the AGN fraction $f_{\rm agn}(M,z)$ can be obtained from the best-fit model in Eq.~\ref{eq:fagn}. The last term of $n_{\rm eagn}^{\rm exp}$ is the volume density of binary AGNs due to random pairing. It needs to be subtracted because each binary AGN is counted only once in the either-AGN sample. 

Because the stellar mass in the NSA catalog is calculated using an elliptical Petrosian aperture that enclose both components of a galaxy pair, we treat it as the total mass and use the mass ratio from the best-fit spectral models (\S~\ref{sec:class}) to split it between the primary and the secondary. 

\begin{figure*}[!tb]
\epsscale{1.2}
\plotone{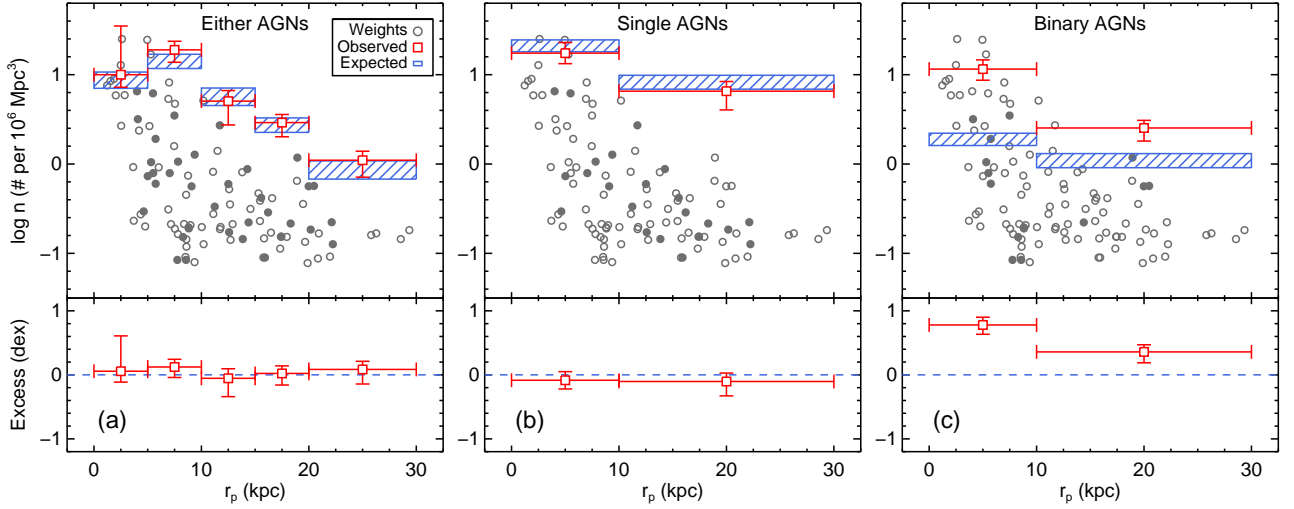}
\caption{Observed vs. expected AGN volume densities (in units of $10^{-6}$~Mpc$^{-3}$) as a function of projected separation for ($a$) either-AGNs, ($b$) single AGNs, and ($c$) binary AGNs. The various AGN samples in paired galaxies are defined in \S~\ref{sec:agn_excess}. In the top panels, the observed volume densities are plotted as red squares with error bars. The horizontal range bar indicates the size of the bin, while the vertical error bar indicates the 1$\sigma$ confidence interval. The blue rectangles show the expected volume densities in each bin (assuming $\xi = 0$ in Eqs.~\ref{eq:neagn_expx}, \ref{eq:nsagn_expx}, and \ref{eq:n2agn_expr}), and their heights indicate the 1$\sigma$ confidence intervals. The gray circles show the individual $1/V_{\rm max}$ weight (Eq.~\ref{eq:weight}) and the {\it gray filled circles} highlight the pairs hosting the relevant AGNs in each sample. The bottom panels show the excess of observed AGNs using the logarithmic ratios of the observed and expected volume densities. Significant excess in volume densities is detected among binary AGNs but not among either-AGNs and single AGNs, indicating that correlated AGN activity is responsible for most of the binary AGNs.
\label{fig:phi_sep}} 
\end{figure*}

In Fig.~\ref{fig:phi_sep}$a$ and $b$, we compare the observed and the expected volume densities in multiple bins of projected separations. We use a bootstrap resampling method to estimate the 1$\sigma$ confidence intervals of the volume densities. The uncertainties of the expected volume densities also includes the 0.04~dex (10\%) error in the best-fit AGN fraction model (Eq.~\ref{eq:fagn}). The bin sizes are chosen as a trade-off between statistical significance and spatial resolution. We find that the observed AGN volume densities are comparable to the expected values at all projected separations for either-AGNs. Interestingly, there is a systematic {\it deficit} of single AGNs at $\sim$1$\sigma$ level at all projected separations, which could be due to correlated AGNs as we will discuss in the following. Therefore, paired galaxies and isolated galaxies have similar AGN duty cycles when matched in stellar mass and redshift, and the mass- and redshift-dependent AGN fractions estimated from the control sample can be applied to paired galaxies.

Then we test whether there is evidence for correlated AGNs by comparing the observed and expected volume densities of binary AGNs. Similar to the previous two AGN samples, the observed volume density of binary AGNs is simply the sum of the $1/V_{\rm max}$ weights: $n_{\rm bagn}^{\rm obs} = \sum_{j=1}^{N_{\rm bagn}} W_j$.
But the expected volume density have two separate parts. One part is from random pairing of AGNs (i.e., uncorrelated): 
\begin{equation} \label{eq:n2agn_expr}
n_{\rm bagn}^{\rm exp,r} = \sum_{j=1}^{N_{\rm pair}} (1-\xi)~W_j~f_{\rm agn}(M_{j}^p,z_j)~f_{\rm agn}(M_{j}^s,z_j)
\end{equation}
The other part is from correlated AGN activity:
\begin{equation} \label{eq:n2agn_expx}
n_{\rm bagn}^{\rm exp,x} = \sum_{j=1}^{N_{\rm pair}} \xi~W_j~[f_{\rm agn}(M_{j}^p,z_j)+f_{\rm agn}(M_{j}^s,z_j)]
\end{equation}
where $\xi$ is the fraction of correlated AGNs. Eq.~\ref{eq:n2agn_expx} gives the volume density of single AGNs that are converted to binary AGNs because of correlated AGN activity. In Fig.~\ref{fig:phi_sep}$c$ we compare the observed volume density with that expected from random pairing with $\xi$ set to zero (Eq.~\ref{eq:n2agn_expr}). Due to the limited number of binary AGNs, we divided the sample into only two bins in projected separation. Contrary to the first two panels, we observe a clear excess of binary AGNs over the expectation from random pairing. The excess is present at all projected separations. Evidently, the excess is due to correlated AGNs and it can be used to estimate the correlated fraction $\xi$ defined in Eq.~\ref{eq:n2agn_expx}. Fig.~\ref{fig:phi_sep}$c$ shows that, as the projected separation decreases from $10-30$~kpc to $1-10$~kpc, the excess of binary AGNs increases from $0.36^{+0.11}_{-0.17}$~dex ($\sim$2.3$\times$) to $0.78^{+0.12}_{-0.14}$~dex ($\sim$6.0$\times$), and correspondingly, $\xi$ increases from $15^{+8}_{-9}$\% to $40^{+16}_{-11}$\%. In spite of different methods, our finding of separation-dependent correlated AGN activity is consistent with previous results based on galaxy pairs identified with single-fiber SDSS spectra \citep{Ellison11,Liu12}. 

Correlated AGNs affect the expected AGN volume densities in galaxy pairs in different ways, depending on how we define the AGN sample:
\begin{enumerate}

\item For either-AGNs, the expected volume density becomes:
\begin{equation} \label{eq:neagn_expx}
\begin{aligned}
n_{\rm eagn}^{\rm exp} &= \sum_{j=1}^{N_{\rm pair}} W_j~f_{\rm agn}(M_{j}^p,z_j)+\sum_{j=1}^{N_{\rm pair}} W_j~f_{\rm agn}(M_{j}^s,z_j) \\ &- \sum_{j=1}^{N_{\rm pair}}(1-\xi)~W_j~f_{\rm agn}(M_{j}^p,z_j)~f_{\rm agn}(M_{j}^s,z_j)\\
\end{aligned}
\end{equation}
The equation shows that correlated AGNs do not significantly affect the population of either-AGNs because one of the AGNs in a correlated binary must be an AGN originally and randomly paired AGNs are the minority among either-AGNs. So the expected volume densities in Fig.~\ref{fig:phi_sep}$a$ remain essentially the same for $\xi \simeq 20-40\%$. 

\item For single AGNs, correlated AGN activity reduces their population by changing a fraction of them to binary AGNs. Given a correlated fraction of $\xi$, we would expect $\xi$ times fewer single AGNs, i.e.,
\begin{equation} \label{eq:nsagn_expx}
\begin{aligned}
n_{\rm sagn}^{\rm exp} &= \sum_{j=1}^{N_{\rm pair}} (1-\xi)~W_j~f_{\rm agn}(M_{j}^p,z_j)~[1-f_{\rm agn}(M_{j}^s,z_j)] \\ &+ \sum_{j=1}^{N_{\rm pair}} (1-\xi)~W_j~f_{\rm agn}(M_{j}^s,z_j)~[1-f_{\rm agn}(M_{j}^p,z_j)] 
\end{aligned}
\end{equation}
This explains the $\sim$0.1~dex (i.e., $\xi \simeq 20\%$) deficit of single AGNs seen in Fig.~\ref{fig:phi_sep}$b$. 

\item The additional population of AGNs produced by correlated activity increases the overall AGN fraction in paired galaxies. When counting all AGNs in paired galaxies (i.e., including the $1/V_{\rm max}$ weights from single AGNs and from {\it all} members of binary AGNs), we detect an AGN excess of $\sim$0.3~dex or $\sim$2$\times$ at $r_p < 30$~kpc, consistent with previous results \citep[e.g.,][]{Ellison11}. This excess can be explained by correlated AGNs, so it does not necessarily suggest a higher AGN duty cycle in paired galaxies.

\end{enumerate}

In summary, assuming the mass-dependent AGN duty cycle from the control sample and a significant fraction of correlated AGNs, our model can explain not only the observed volume densities of either-AGNs, but also the observed deficit of single AGNs and the excess of binary AGNs. On the other hand, there is no evidence for enhanced AGN duty cycles in paired galaxies other than the additional population of AGNs produced by correlated activities. An alternative model assuming significantly higher AGN duty cycles in paired galaxies than in the control sample cannot explain the observed volume densities of the various subsamples simultaneously. Therefore, we conclude that correlated AGN activities are significant in interacting pairs, at least when the projected separations are less than 30~kpc and they account for the majority of the observed binary AGNs. In the next section, we discuss the possible mechanisms that could have produced correlated AGNs.

\section{Summary and Discussion} \label{sec:summary}

Based on the SDSS-IV MaNGA integral-field spectroscopic data of 2,618 galaxies at $0.01 < z < 0.15$, we have identified a sample of 105 kinematic galaxy pairs with projected separations less than 30~kpc, radial velocity offsets less than 600~\kms, and mass ratios greater than 0.1. Based on this sample and a careful treatment of selection biases in both the pair sample and the MaNGA parent sample, we measure the fraction of galaxy pairs in the nearby universe and compare the incidence of AGNs in paired galaxies and isolated galaxies. We summarize our main findings below:

\begin{enumerate}

\item The pair fraction increases almost linearly with the physical size of the IFU and the logarithmic of the stellar mass, consistent with theoretical expectations. Consequently, the pair sample is biased to higher stellar masses and higher redshifts, when compared to the MaNGA sample.

\item Integrating the pair fraction over a volume-limited sample, we find that $3.4\pm0.5$\% of $M^*$ galaxies at $\bar{z} = 0.04$ are in major- and minor-merger galaxy pairs with projected separations less than $20h^{-1}$~kpc = 28.6~kpc. 

\item Among isolated galaxies, AGNs selected using the \citet{Kauffmann03} dividing line and the H$\alpha$ EW cutoff at 3~\AA\ are strongly biased to host galaxies with stellar mass around $4\times10^{10}$~\msun. The stellar-mass bias can be adequately modeled with a broad Gaussian function with a width of $\sigma = 0.54$~dex.

\item The apparently high AGN fraction in paired galaxies is primarily due to the selection biases of the pair sample. Discounting correlated AGNs, the observed volume densities of AGNs in paired galaxies agree well with the expectation from the observed AGN fraction in isolated galaxies at all projected separations, suggesting that interactions do not increase the AGN duty cycle. 

\item There are 14 binary AGNs, accounting for $\sim$13\% of the pair sample. The incidence of binary AGNs cannot be explained by random pairing of AGNs, given our finding that paired galaxies and isolated galaxies have similar AGN duty cycles. Instead, about 60-80\% of the binary AGNs are due to correlated activities, where an AGN makes its companion galaxy also an AGN. The correlated fraction increases as the projected separation decreases.

\end{enumerate}

Our results highlight the importance of correlated AGNs in close galaxy pairs. What could have caused a correlation of AGNs in a pair of galaxies? We consider three possible channels that could produce narrow-line-selected binary AGNs: (1) synchronized SMBH accretion due to tidal interactions, (2) wide-spread radiative shocks driven by gas flows induced in galaxy interactions, and (3) AGN cross-ionization of companion galaxies. 

First, numerical simulations of galaxy mergers suggest that merger-triggered AGNs may be important in producing binary AGNs \citep[e.g.,][]{Van-Wassenhove12}. This is because the sudden gas inflow driven by tidally-induced stellar bars could be an important triggering mechanism for AGNs and the periods when each galaxy is an AGN may have significant overlap, leading to an episode of synchronized SMBH accretion. However, suppose a significant fraction of AGNs are triggered by tidally-induced gas inflows, one would expect higher AGN duty cycles in paired galaxies than in isolated galaxies. This is inconsistent with the observed volume densities of the three AGN subsamples we calculated in Fig.~\ref{fig:phi_sep}. For example, adopting an AGN duty cycle significantly higher than the control sample would increase the expected volume densities of either-AGNs above the observed values, leading to a deficit that is difficult to explain. On the other hand, the majority of the low-luminosity AGNs are not in mergers (as we find in MaNGA), indicating the importance of minor disturbances in triggering AGNs.

Secondly, the AGN branch in the [N\,{\sc ii}]/H$\alpha$ BPT diagram represents mixing sequences of gas photoionized by star formation and more energetic ionization sources such as AGNs, evolved stellar populations \citep{Stasinska08}, or radiative shocks \citep{Allen08}. Gas predominantly photoionized by evolved stellar populations can be removed by applying an empirical cut in H$\alpha$ EW (as we did in \S~\ref{sec:class}). It is more difficult to separate AGNs from radiative shocks. In merging galaxies, galactic-scale shocks could be produced by gas infall driven by tidally-induced stellar bars \citep{Monreal-Ibero10,Rich15}. Observationally, merger-driven shocks seem prevalent in local ultra-luminous and luminous infrared galaxies (U/LIRGs), especially in late-stage mergers when the nuclei are separated less than 10~kpc or have coalesced \citep{Rich15}. Given that $\sim$40\% of local LIRGs and $\sim$55\% of ULIRGs in close binary phase show starburst-AGN composite nuclear spectra \citep{Yuan10}, radiative shocks could be an important mechanism to produce such ``apparent'' AGNs. By cross-matching with the IRAS Revised Bright Galaxy Sample \citep{Sanders03}, we find two LIRGs in the MaNGA sample: 7443-12703 (Mrk~848) and 8250-12704 (Arp~55). Both LIRGs are close binaries with $r_p < 10$~kpc (i.e., interaction class IIIb; \citealt{Veilleux02a}) and both show wide-spread starburst-AGN composite emission-line ratios. The former is classified as a binary AGN, while the latter is a SF-Composite pair, suggesting that at least one of our binary AGNs could be due to merger-driven shocks instead of active SMBH accretion. Merger-driven shocks naturally cause correlated line ratios between merging components, because gas infall occurs in both galaxies almost simultaneously. To explain the similar AGN duty cycles in paired and isolated galaxies, it would require that shock-ionization are equally important in isolated galaxies and interacting galaxies. This is possible because shocks in isolated galaxies could be driven by galactic-scale outflows from nuclear starbursts \citep{Rich10,Sharp10,Ho14}. 

Last but not the least, the UV and X-ray photons produced by an AGN can ionize gas at great distances. Quasars show extended emission-line regions (EELRs) that stretch tens of kpc from the SMBHs \citep{Stockton06}, and the size of the EELRs slowly decreases with the AGN luminosity \citep[e.g.,][]{Greene11,Fu12a}, similar to the Str\"omgren sphere of an H\,{\sc ii} region. Therefore, the interstellar gas in a galaxy can be photoionized by a nearby AGN instead of an AGN in the galaxy itself \citep{Keel17a}. When cross-ionization occurs, a galaxy pair would appear to be a binary AGN even though only one of the members is an AGN. Because cross-ionization requires the companion galaxy within the AGN ionization cone, the fraction of correlated AGNs ($\xi$) sets an upper limit on the solid angle opening of the ionization cone: $\Omega/2\pi < \xi$. The correlated fraction decreases with projected separation because more luminous AGNs are needed to cross-ionize companion galaxies at greater distances.

Therefore, merger-driven shocks and AGN cross-ionization could be the primary channels of producing correlated binary AGNs. But in neither channel are there two {\it bona fide} AGNs (i.e., accreting SMBHs) in a system, so these should be considered as ``apparent'' binary AGNs. Future studies could confirm/refute the two channels with better diagnostics or avoid such ``apparent'' binaries by selecting AGNs in less ambiguous wavelengths (e.g., X-ray, mid-IR, or radio). 

\acknowledgments

We thank the anonymous referee for cogent comments that helped improve the presentation of the paper. We thank Mike Boylan-Kolchin for helpful discussions on dynamical friction. H.F., J.W.I., and J.L.S. acknowledge support from the National Science Foundation (NSF) grant AST-1614326. This work made use of the JBIU IDL library, available at \url{http://www.simulated-galaxies.ua.edu/jbiu/}.
Funding for the Sloan Digital Sky Survey IV has been provided by
the Alfred P. Sloan Foundation, the U.S. Department of Energy Office of
Science, and the Participating Institutions. SDSS-IV acknowledges
support and resources from the Center for High-Performance Computing at
the University of Utah. The SDSS web site is www.sdss.org.
SDSS-IV is managed by the Astrophysical Research Consortium for the 
Participating Institutions of the SDSS Collaboration including the 
Brazilian Participation Group, the Carnegie Institution for Science, 
Carnegie Mellon University, the Chilean Participation Group, the French Participation Group, Harvard-Smithsonian Center for Astrophysics, 
Instituto de Astrof\'isica de Canarias, The Johns Hopkins University, 
Kavli Institute for the Physics and Mathematics of the Universe (IPMU) / 
University of Tokyo, Lawrence Berkeley National Laboratory, 
Leibniz Institut f\"ur Astrophysik Potsdam (AIP),  
Max-Planck-Institut f\"ur Astronomie (MPIA Heidelberg), 
Max-Planck-Institut f\"ur Astrophysik (MPA Garching), 
Max-Planck-Institut f\"ur Extraterrestrische Physik (MPE), 
National Astronomical Observatory of China, New Mexico State University, 
New York University, University of Notre Dame, 
Observat\'ario Nacional / MCTI, The Ohio State University, 
Pennsylvania State University, Shanghai Astronomical Observatory, 
United Kingdom Participation Group,
Universidad Nacional Aut\'onoma de M\'exico, University of Arizona, 
University of Colorado Boulder, University of Oxford, University of Portsmouth, 
University of Utah, University of Virginia, University of Washington, University of Wisconsin, 
Vanderbilt University, and Yale University.

\bibliographystyle{apj}
\bibliography{exgal_ref}

\clearpage

\appendix

\section{MaNGA Binary AGNs and Pair Catalog}

Fig.~\ref{fig:bpt2d} shows the SDSS pseudo-color images and the spatially resolved gas kinematics, emission-line ratios, and BPT diagnostic maps obtained from fitting the full datacubes. Tidal features are evident in the SDSS images for many of the binary AGNs, indicating ongoing interactions. The DRP-produced datacubes have been rebinned from the original 0.5\arcsec\ spaxels to 1\arcsec\ spaxels to remove much of the covariances between spectra in neighboring spaxels. Only spaxels with H$\alpha$ A/N~$>$~3 are shown in the figure because the emission-line ratios are much less reliable at lower A/Ns. The sample displays complex gas kinematics, and a variety of line-ratio topography and mixing sequences in the BPT diagram.

Table~\ref{tab:pairs} lists the 105 galaxy pairs selected from the MaNGA main galaxy sample (\S~\ref{sec:pair_selection}). We list the 14 binary AGNs separately from the rest of the sample.

\begin{figure*}[!h]
\epsscale{1.2}
\plotone{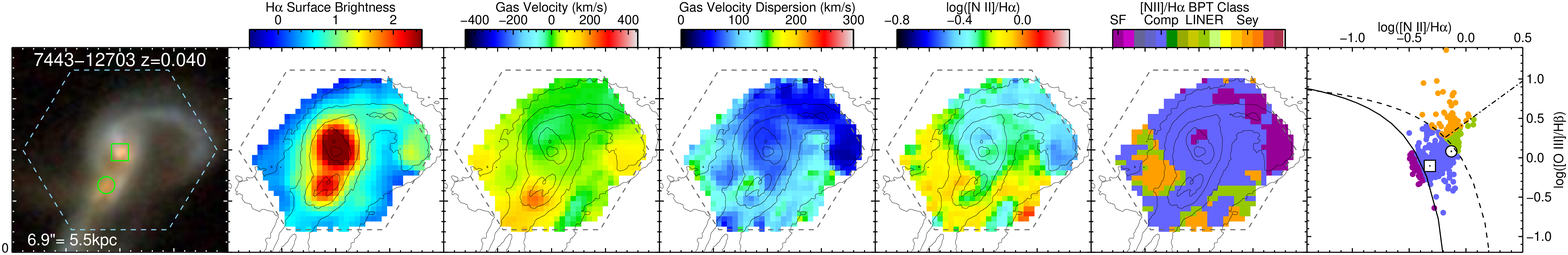}
\plotone{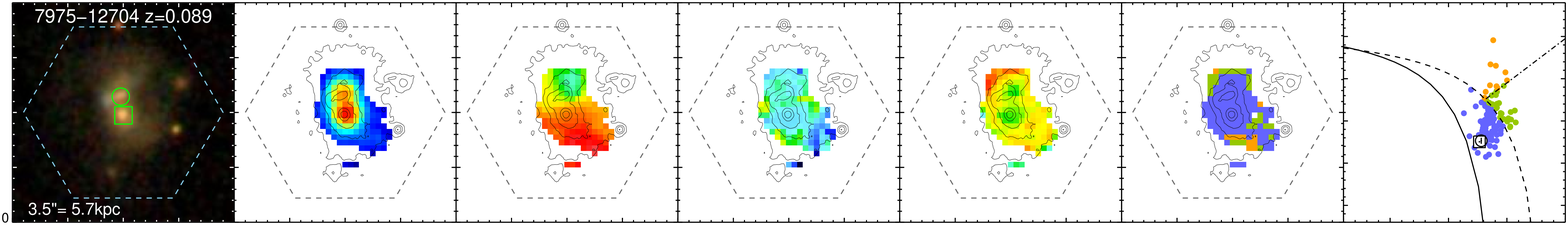}
\plotone{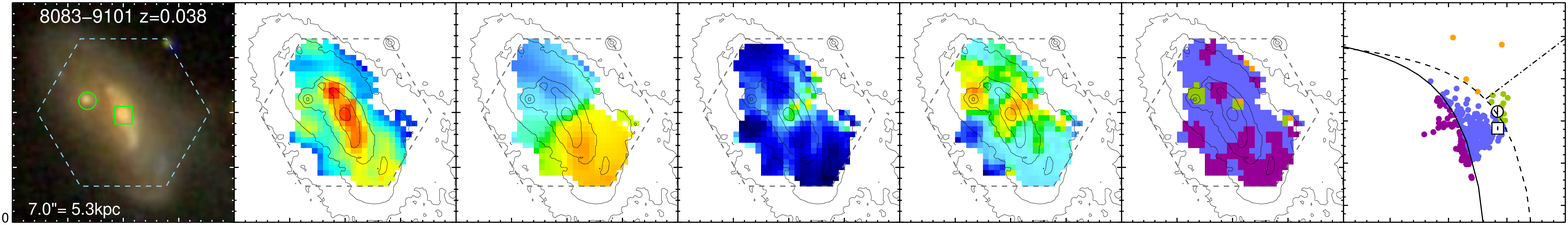}
\plotone{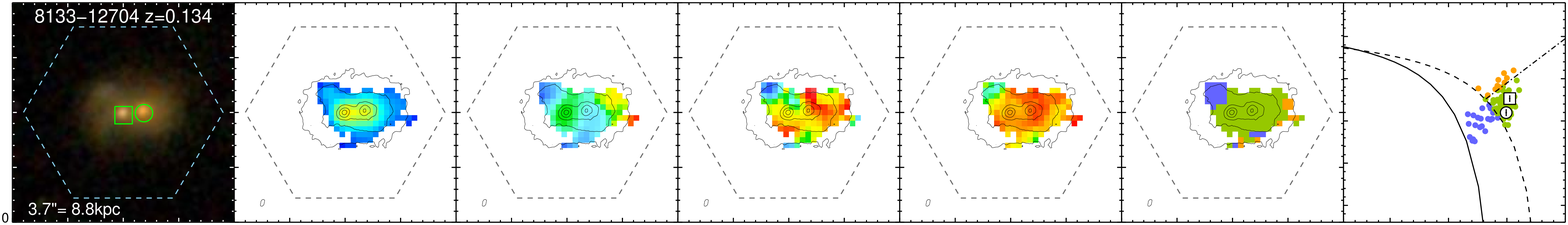}
\plotone{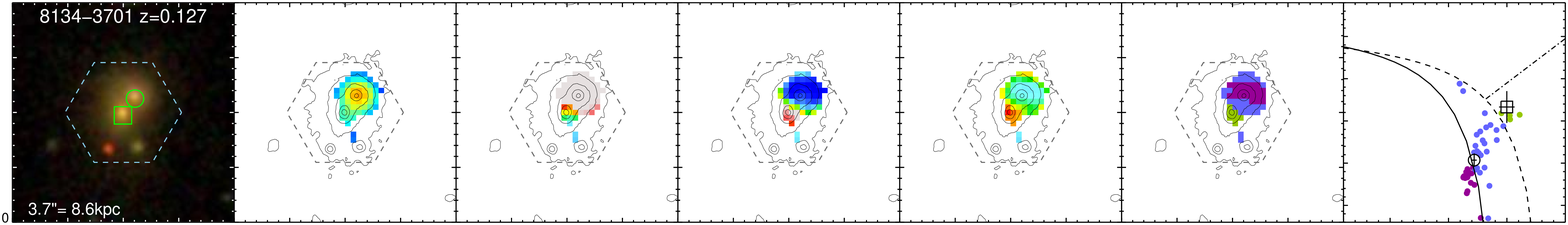}
\plotone{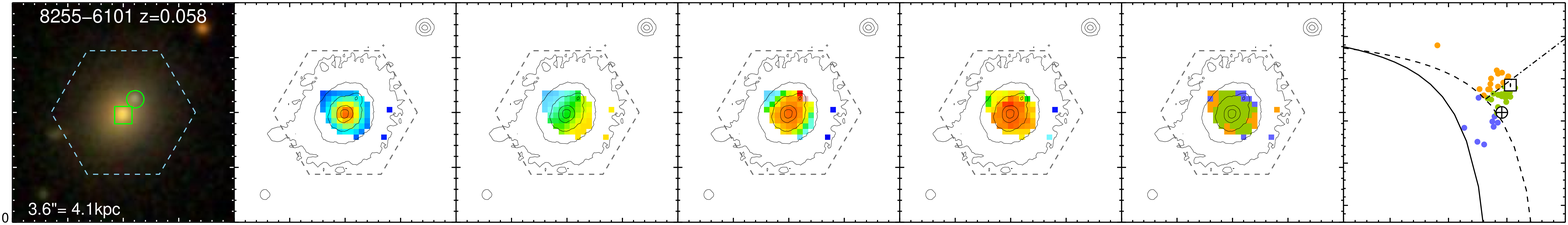}
\caption{Binary AGNs in SDSS-IV MaNGA. For each binary, we show from left to right, the SDSS $gri$ pseudo-color image, maps of H$\alpha$ emission surface brightness (in units of $10^{-17}$~erg~s$^{-1}$~cm$^{-2}$~arcsec$^{-2}$), gas velocity, gas velocity dispersion, [N\,{\sc ii}]/H$\alpha$ line ratio, [N\,{\sc ii}]/H$\alpha$ BPT classification, and the distribution of the spaxels on the [N\,{\sc ii}]/H$\alpha$ BPT diagram. We only show spaxels with H$\alpha$ amplitude-to-noise ratio greater than 3. The MaNGA ID, the redshift, and the projected angular and physical separations are labeled. The BPT classification map and the BPT diagram share the same color code and the demarcation lines in the latter are the same as those in Fig.~\ref{fig:bptwhan}$a$. In the first and the last columns, we indicate the locations of the primary and the secondary components of each pair with a square and a circle, respectively. All maps are 40\arcsec\ across, with minor tickmarks spaced in 2\arcsec\ intervals. The dashed hexagons show the IFU FoV. The contours are from the SDSS pseudo-color image, drawn at (0.15, 0.34, 0.53, 0.72, 0.9)$\times$ the maximum pixel value. 
\label{fig:bpt2d}} 
\end{figure*}
\addtocounter{figure}{-1}
\begin{figure*}
\epsscale{1.2}
\plotone{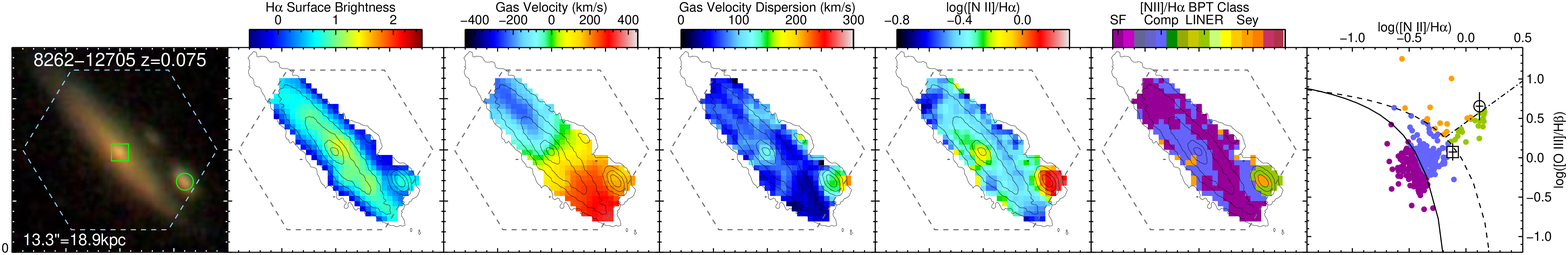}
\plotone{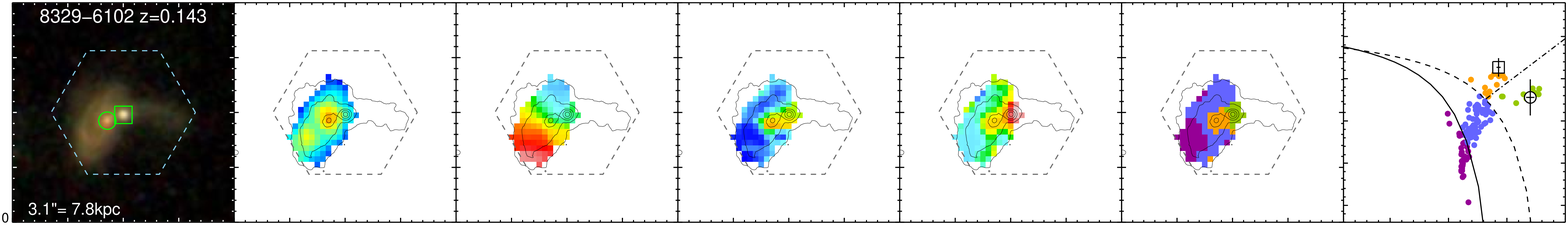}
\plotone{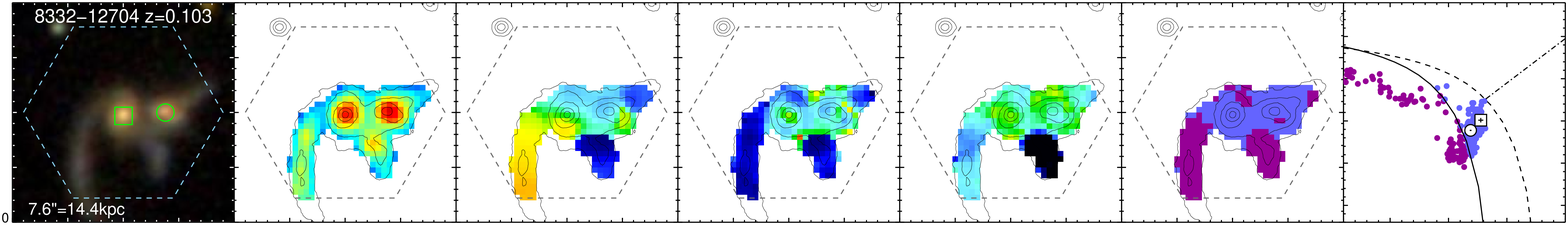}
\plotone{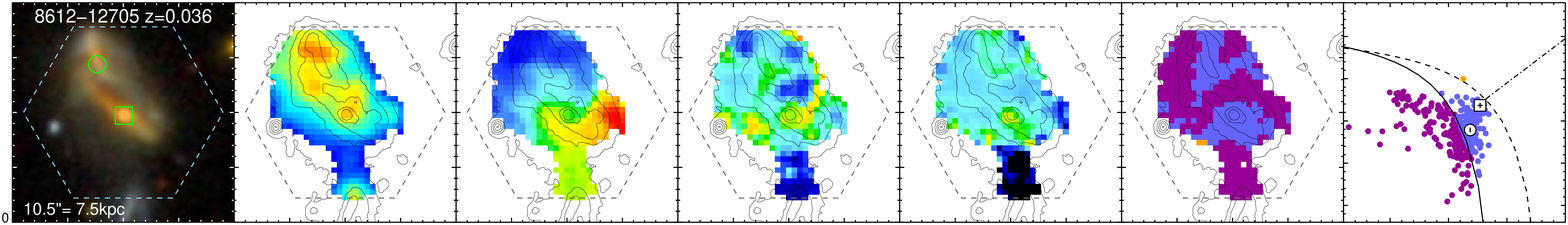}
\plotone{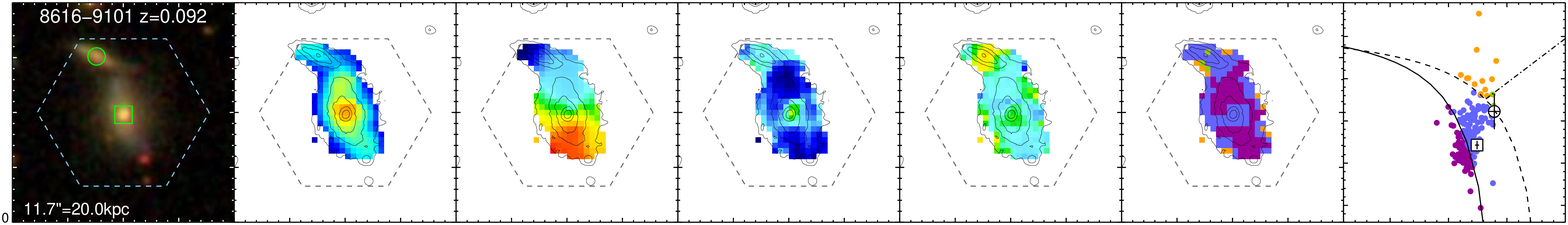}
\plotone{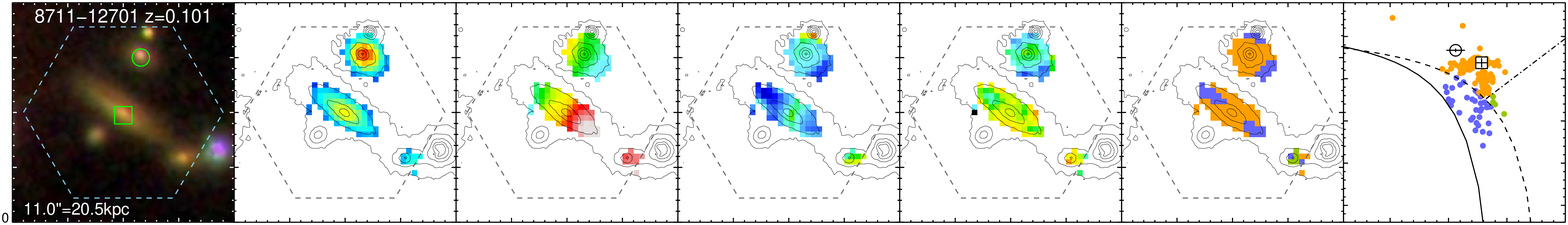}
\plotone{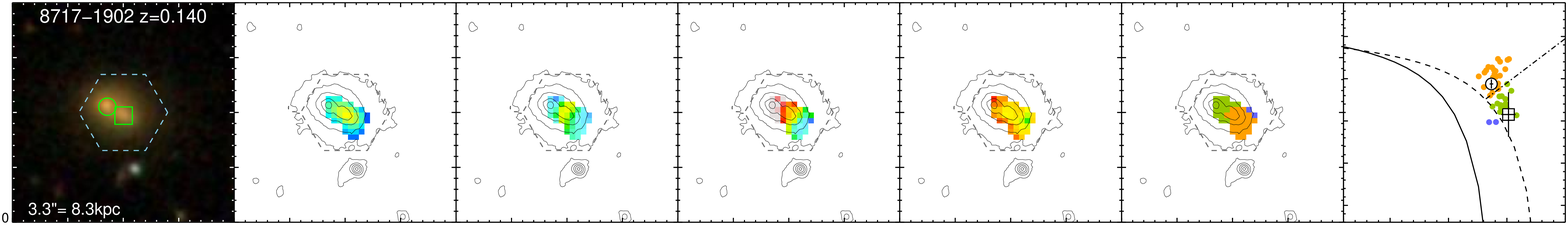}
\plotone{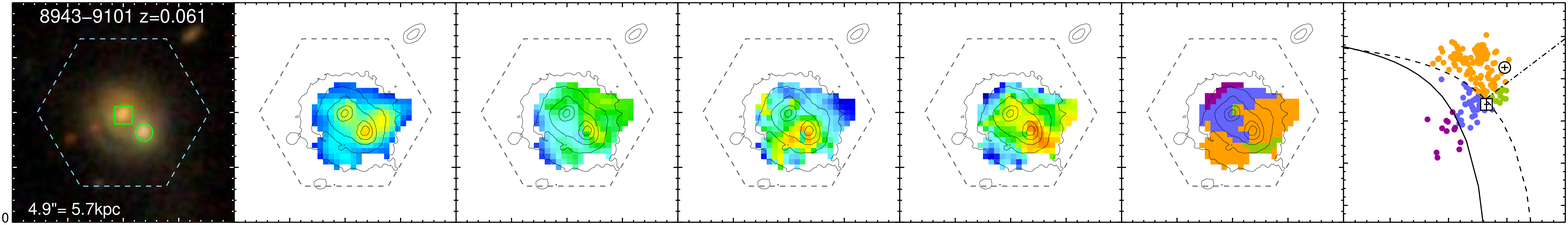}
\caption{{\it continued}.}
\end{figure*}

\clearpage
\LongTables
\begin{deluxetable*}{rrcrrrrrrrrcc}
\tablewidth{0pt}
\tablecaption{Galaxy Pairs in SDSS-IV MaNGA (DR14).
\label{tab:pairs}}
\tablehead{
\colhead{Plate} & \colhead{IFU} & \colhead{Sample} & \colhead{$\Delta\theta$} & \colhead{$r_p$} & \colhead{$\Delta v$} & \colhead{R.A.} & \colhead{Decl.} & \colhead{Redshift} & \colhead{$\sigma_{\rm star}$} & \colhead{log($M$)} & \colhead{log($L_{\rm [OIII]}$)} & \colhead{Class} \\
\colhead{} & \colhead{} & \colhead{} & \colhead{arcsec} & \colhead{kpc} & \colhead{\kms} & \colhead{deg} & \colhead{deg} & \colhead{} & \colhead{\kms} & \colhead{log(\msun)} & \colhead{log(\ergs)} & \colhead{} \\
\colhead{(1)} & \colhead{(2)} & \colhead{(3)} & \colhead{(4)} & \colhead{(5)} & \colhead{(6)} & \colhead{(7)} & \colhead{(8)} & \colhead{(9)} & \colhead{(10)} & \colhead{(11)} & \colhead{(12)} & \colhead{(13)} 
}
\startdata
\multicolumn{13}{c}{Binary AGNs}\\	 	 	 	            
\hline	 
7443&12703&1&  6.9&  5.5& 91.9&229.52558&$+$42.74585&0.04044& 205.1&   9.2&  40.6&2\\
    &     & &     &     &     &229.52653&$+$42.74407&0.04074& 126.0&   9.3&  39.9&2\\
7975&12704&2&  3.5&  5.7&178.6&324.58641&$+$11.34867&0.08961& 160.3&   9.8&  39.8&2\\
    &     & &     &     &     &324.58655&$+$11.34961&0.08902& 154.7&   9.1&  39.3&2\\
8083& 9101&1&  7.0&  5.3&122.4& 50.13841& $-$0.33996&0.03828& 174.0&  10.1&  39.5&2\\
    &     & &     &     &     & 50.14021& $-$0.33923&0.03788&  62.2&   9.3&  39.1&3\\
8133&12704&2&  3.7&  8.8& 36.9&114.77573&$+$44.40277&0.13425& 246.6&   9.7&  39.6&3\\
    &     & &     &     &     &114.77430&$+$44.40287&0.13437& 177.6&   9.3&  39.4&3\\
8134& 3701&3&  3.7&  8.6&547.6&113.32663&$+$46.34819&0.12719& 252.0&   9.9&  39.0&3\\
    &     & &     &     &     &113.32576&$+$46.34904&0.12902& 233.0&   9.5&  39.0&2\\
8255& 6101&2&  3.6&  4.1&  7.5&166.50988&$+$43.17348&0.05841& 191.8&  10.1&  40.0&3\\
    &     & &     &     &     &166.50906&$+$43.17428&0.05843& 197.1&   9.4&  39.0&3\\
8262&12705&2& 13.3& 18.9&127.5&185.49815&$+$44.14566&0.07486& 137.8&   9.6&  39.0&2\\
    &     & &     &     &     &185.49350&$+$44.14407&0.07529&  95.2&   9.1&  39.0&4\\
8329& 6102&3&  3.1&  7.8& 76.8&211.90486&$+$44.48229&0.14310& 213.3&   9.6&  40.3&4\\
    &     & &     &     &     &211.90599&$+$44.48199&0.14285& 140.9&   9.4&  39.2&3\\
8332&12704&2&  7.6& 14.4& 78.3&209.16355&$+$43.58561&0.10308& 152.0&   9.7&  40.0&2\\
    &     & &     &     &     &209.16066&$+$43.58578&0.10282& 140.0&   8.9&  39.9&2\\
8612&12705&3& 10.5&  7.5& 50.1&255.10152&$+$38.35170&0.03572& 136.2&   9.8&  39.5&2\\
    &     & &     &     &     &255.10322&$+$38.35430&0.03556& 117.6&   9.3&  39.2&2\\
8616& 9101&2& 11.7& 20.0&239.1&322.90884& $-$0.60818&0.09167& 192.8&   9.8&  39.2&2\\
    &     & &     &     &     &322.91019& $-$0.60523&0.09087& 115.3&   9.2&  38.6&2\\
8711&12701&2& 11.0& 20.5&214.7&116.94308&$+$51.64602&0.10169& 177.4&   9.5&  39.8&4\\
    &     & &     &     &     &116.94167&$+$51.64894&0.10097& 173.1&   9.3&  40.6&5\\
8717& 1902&2&  3.3&  8.3& 29.6&118.09109&$+$34.32655&0.13989& 245.7&   9.8&  39.1&3\\
    &     & &     &     &     &118.09207&$+$34.32700&0.13999& 217.5&   9.6&  39.7&4\\
8943& 9101&2&  4.9&  5.7& 35.3&156.40312&$+$37.22230&0.06064& 125.5&   9.8&  39.3&2\\
    &     & &     &     &     &156.40182&$+$37.22143&0.06076& 162.9&   9.4&  40.0&4\\
\hline	 	 	 	            
\multicolumn{13}{c}{Remaining Pairs}\\	 	 	 	            
\hline	 
7495&12703&1& 11.6&  6.9&112.0&206.47068&$+$26.77517&0.02983&  71.0&   9.4&  38.6&1\\
    &     & &     &     &     &206.46873&$+$26.77247&0.02945&  59.8&   8.6&  39.5&1\\
7975& 6104&2&  4.7&  7.1&  7.2&324.89157&$+$10.48348&0.07887& 111.2&   9.8&  39.0&1\\
    &     & &     &     &     &324.89036&$+$10.48405&0.07885& 118.4&   9.3&  38.5&2\\
7975&12702&2&  6.4&  9.4&213.7&323.52119&$+$10.42183&0.07753& 117.6&   9.4&  39.3&2\\
    &     & &     &     &     &323.52173&$+$10.42352&0.07682&  71.1&   8.6&  40.1&1\\
7991&12702&1&  5.7&  3.7&213.1&258.84693&$+$57.43288&0.03224& 212.9&  10.4&  \nod&0\\
    &     & &     &     &     &258.84750&$+$57.43133&0.03153& 127.9&  10.0&  \nod&0\\
7992&12704&1& 11.3& 20.8&371.3&254.71733&$+$63.17955&0.09878& 272.9&  10.4&  \nod&0\\
    &     & &     &     &     &254.71588&$+$63.17648&0.10001& 154.7&   9.9&  \nod&0\\
8086& 6101&2&  8.4& 19.6&465.5& 57.49040& $+$0.07435&0.13158& 237.7&  10.1&  \nod&0\\
    &     & &     &     &     & 57.48843& $+$0.07558&0.13313& 168.4&   9.4&  \nod&0\\
8131&12702&1& 11.0& 15.5&460.0&112.36028&$+$40.16919&0.07356& 271.7&  10.4&  \nod&0\\
    &     & &     &     &     &112.36356&$+$40.17097&0.07203& 194.9&  10.0&  \nod&0\\
8132&12701&2& 10.7& 25.8&433.2&110.51721&$+$42.42137&0.13571& 288.3&  10.1&  \nod&0\\
    &     & &     &     &     &110.51661&$+$42.42431&0.13716& 204.2&   9.7&  \nod&0\\
8138& 1901&3&  3.4&  1.9& 73.7&116.76382&$+$46.53450&0.02742&  65.8&   8.7&  39.3&1\\
    &     & &     &     &     &116.76253&$+$46.53480&0.02767&  75.8&   8.6&  39.5&1\\
8140& 3701&3&  3.3&  2.5& 57.3&116.17684&$+$42.35765&0.03911&  81.6&   9.2&  39.2&1\\
    &     & &     &     &     &116.17627&$+$42.35684&0.03930& 104.2&   8.9&  39.5&1\\
8140& 3704&2&  7.9& 11.5& 78.8&116.78570&$+$40.39294&0.07647& 188.3&  10.0&  \nod&0\\
    &     & &     &     &     &116.78388&$+$40.39123&0.07621&  68.4&   9.1&  \nod&0\\
8143&12704&1&  8.2&  5.2&  3.2&119.55999&$+$42.26865&0.03177& 121.9&   9.9&  \nod&0\\
    &     & &     &     &     &119.56306&$+$42.26869&0.03176&  79.4&   9.4&  39.1&1\\
8145& 3702&1&  7.8&  2.5&  1.7&116.37954&$+$28.44099&0.01576&  61.5&   8.5&  40.0&1\\
    &     & &     &     &     &116.37714&$+$28.44056&0.01576&  44.8&   7.9&  38.3&1\\
8146& 3702&1&  5.4&  4.2& 40.1&116.11040&$+$29.26940&0.03953& 189.7&  10.4&  \nod&0\\
    &     & &     &     &     &116.10886&$+$29.26876&0.03967&  87.7&   9.5&  \nod&0\\
8146&12704&1& 12.0& 14.6&519.6&117.48112&$+$29.42019&0.06282& 208.9&  10.2&  \nod&0\\
    &     & &     &     &     &117.48326&$+$29.41742&0.06456& 132.0&   9.7&  \nod&0\\
8149& 9102&1&  2.8&  1.2& 49.1&120.97918&$+$26.52062&0.02172&  64.3&   9.3&  39.2&1\\
    &     & &     &     &     &120.97984&$+$26.52011&0.02189&  48.8&   8.9&  39.2&1\\
8156& 6103&2&  4.8& 11.8&226.1& 54.52662& $+$0.82449&0.13901& 231.0&  10.0&  \nod&0\\
    &     & &     &     &     & 54.52629& $+$0.82578&0.13977& 134.5&   9.1&  \nod&0\\
8156& 9101&3&  6.4& 12.7& 52.7& 56.42513& $-$0.37844&0.10936& 216.3&   9.9&  \nod&0\\
    &     & &     &     &     & 56.42661& $-$0.37942&0.10919& 153.4&   9.7&  \nod&0\\
8239& 6103&3&  9.1& 16.5&235.5&115.67203&$+$48.82605&0.09724& 282.9&  10.3&  \nod&0\\
    &     & &     &     &     &115.67426&$+$48.82398&0.09646& 141.0&   9.6&  \nod&0\\
8239&12701&2&  3.1&  7.2& 87.9&115.38745&$+$47.87100&0.12760& 212.4&   9.9&  \nod&0\\
    &     & &     &     &     &115.38750&$+$47.87187&0.12730& 294.0&   9.7&  \nod&0\\
8241& 6103&1&  4.6&  8.6& 87.4&126.13716&$+$19.26494&0.10008& 205.7&   9.4&  \nod&0\\
    &     & &     &     &     &126.13847&$+$19.26529&0.10038& 267.4&  10.3&  \nod&0\\
8244& 1901&1&  5.4&  9.0&203.9&132.54470&$+$51.77540&0.08878& 236.0&  10.3&  \nod&0\\
    &     & &     &     &     &132.54499&$+$51.77689&0.08810& 165.8&   9.8&  \nod&0\\
8250&12701&1& 12.4& 20.2&134.5&139.19315&$+$42.50677&0.08659& 229.9&  10.2&  \nod&0\\
    &     & &     &     &     &139.18995&$+$42.50929&0.08614& 157.1&   9.5&  39.0&2\\
8250&12704&1& 11.5&  9.1&108.8&138.98134&$+$44.33280&0.03988& 150.4&   9.4&  39.4&2\\
    &     & &     &     &     &138.97780&$+$44.33084&0.03951& 154.2&   9.1&  39.4&1\\
8253&12702&1& 12.2& 22.2&248.7&157.62416&$+$43.24222&0.09875& 220.6&  10.0&  39.1&3\\
    &     & &     &     &     &157.62824&$+$43.24381&0.09958& 244.8&  10.1&  \nod&0\\
8253&12705&2& 11.5& 18.9&315.5&158.78531&$+$42.63880&0.08667& 214.7&   9.8&  \nod&0\\
    &     & &     &     &     &158.78103&$+$42.63828&0.08772&  80.5&   9.4&  38.5&1\\
8256& 9101&1&  6.2&  6.0&565.1&164.97312&$+$43.36975&0.04960& 249.3&  10.5&  \nod&0\\
    &     & &     &     &     &164.97109&$+$43.36885&0.04771& 208.4&  10.3&  \nod&0\\
8256&12704&1&  7.0& 15.9&289.0&166.12939&$+$42.62453&0.12601& 227.2&  10.1&  40.7&4\\
    &     & &     &     &     &166.12950&$+$42.62258&0.12505& 232.8&  10.1&  \nod&0\\
8257& 6104&1&  4.0& 10.0&452.7&166.10685&$+$45.78171&0.14284& 280.2&  10.1&  \nod&0\\
    &     & &     &     &     &166.10728&$+$45.78278&0.14435& 264.5&  10.0&  \nod&0\\
8313& 1901&1&  4.1&  2.0&117.2&240.28712&$+$41.88077&0.02422&  87.6&   9.0&  40.9&1\\
    &     & &     &     &     &240.28850&$+$41.88128&0.02461& 800.0&   8.6&  41.3&1\\
8315&12701&2&  6.1&  7.5&208.2&234.55398&$+$38.17420&0.06403& 137.5&  10.0&  \nod&0\\
    &     & &     &     &     &234.55458&$+$38.17258&0.06473& 124.4&   9.9&  \nod&0\\
8317&12701&1&  7.2& 10.2&397.1&192.26991&$+$43.83319&0.07422& 273.4&  10.4&  \nod&0\\
    &     & &     &     &     &192.26757&$+$43.83427&0.07289& 177.4&   9.6&  \nod&0\\
8317&12702&2&  8.0& 11.1&180.6&193.43337&$+$43.24986&0.07251& 203.9&  10.2&  \nod&0\\
    &     & &     &     &     &193.43451&$+$43.25193&0.07191& 118.7&   9.6&  \nod&0\\
8319& 9102&1&  4.6&  7.8& 30.7&201.56098&$+$47.30030&0.09024& 234.1&  10.1&  \nod&0\\
    &     & &     &     &     &201.55958&$+$47.29943&0.09034& 265.1&   9.3&  \nod&0\\
8329& 9101&2&  9.8& 19.5& 70.9&211.25306&$+$44.75245&0.10796& 163.4&   9.8&  \nod&0\\
    &     & &     &     &     &211.24936&$+$44.75319&0.10772& 136.2&   9.6&  \nod&0\\
8329&12705&2&  3.5&  8.8& 16.0&214.54773&$+$44.47426&0.14158& 244.0&   9.9&  \nod&0\\
    &     & &     &     &     &214.54637&$+$44.47439&0.14152& 175.3&   9.4&  \nod&0\\
8330&12702&2& 11.4& 14.3&148.3&203.85305&$+$38.09515&0.06546&  80.0&   9.5&  39.4&1\\
    &     & &     &     &     &203.85572&$+$38.09276&0.06497&  89.7&   9.5&  39.1&2\\
8333&12701&1&  7.5& 12.6&241.1&213.43219&$+$43.66251&0.08931& 244.2&  10.0&  39.7&3\\
    &     & &     &     &     &213.43364&$+$43.66432&0.08851& 136.9&   9.4&  \nod&0\\
8440& 6101&2&  8.1& 20.9& 17.2&134.67459&$+$41.23641&0.14743& 204.7&   9.8&  \nod&0\\
    &     & &     &     &     &134.67200&$+$41.23528&0.14737& 150.0&   8.9&  \nod&0\\
8447& 6101&3&  3.2&  4.6& 57.0&206.13326&$+$40.24002&0.07529& 198.3&  10.4&  \nod&0\\
    &     & &     &     &     &206.13403&$+$40.24070&0.07548& 146.2&   9.7&  38.8&2\\
8447& 9102&1&  9.8& 17.4&244.8&207.45444&$+$40.53743&0.09700& 215.9&   9.9&  39.2&1\\
    &     & &     &     &     &207.45197&$+$40.53938&0.09619& 221.5&   9.9&  39.4&2\\
8448& 3703&3&  3.8&  9.4&101.9&167.08980&$+$22.66146&0.13859& 198.1&   9.6&  38.8&1\\
    &     & &     &     &     &167.09089&$+$22.66178&0.13893& 161.9&   9.6&  38.9&1\\
8454& 6102&2&  7.8& 16.2&115.8&153.53547&$+$44.17574&0.11360& 217.2&   9.8&  \nod&3\\
    &     & &     &     &     &153.53280&$+$44.17676&0.11321& 144.0&   9.5&  \nod&0\\
8454& 9101&2& 12.0& 29.4&135.0&153.92384&$+$45.36708&0.13764& 224.4&   9.8&  \nod&0\\
    &     & &     &     &     &153.92280&$+$45.36382&0.13809&  84.7&   9.0&  \nod&0\\
8458&12703&2& 10.0& 13.8&226.9&148.89778&$+$45.55478&0.07248& 175.6&   9.5&  38.7&1\\
    &     & &     &     &     &148.90088&$+$45.55651&0.07323& 103.8&   9.7&  \nod&0\\
8461&12701&3& 14.3& 19.9& 72.8&146.47820&$+$43.04671&0.07283& 251.9&  10.3&  \nod&0\\
    &     & &     &     &     &146.48319&$+$43.04512&0.07307& 168.2&  10.1&  \nod&0\\
8461&12702&1& 14.1&  5.0&110.2&145.72411&$+$42.79317&0.01720&  47.5&   8.2&  38.6&1\\
    &     & &     &     &     &145.72703&$+$42.79645&0.01757& 586.0&   7.3&  39.6&1\\
8464& 9101&1&  8.6& 17.4&132.5&186.67495&$+$43.78232&0.11156& 222.6&  10.1&  \nod&0\\
    &     & &     &     &     &186.67636&$+$43.78447&0.11200& 194.2&   9.9&  \nod&0\\
8465& 6101&1&  6.2& 15.8&173.6&195.52294&$+$47.49698&0.14348& 287.9&  10.1&  \nod&0\\
    &     & &     &     &     &195.52082&$+$47.49795&0.14406& 190.4&   9.4&  \nod&0\\
8481& 3704&2&  6.3& 12.7&113.7&240.18671&$+$53.77554&0.11012& 269.6&  10.0&  \nod&0\\
    &     & &     &     &     &240.18967&$+$53.77564&0.11050& 201.8&   9.1&  \nod&0\\
8481& 6103&1&  8.5&  7.8&285.2&238.28611&$+$54.14727&0.04673& 135.6&  10.0&  39.2&2\\
    &     & &     &     &     &238.28900&$+$54.14892&0.04768&  92.7&   9.6&  \nod&0\\
8483& 6102&1&  3.9&  4.3&374.5&246.11867&$+$48.59665&0.05697& 251.0&  10.5&  \nod&0\\
    &     & &     &     &     &246.11865&$+$48.59557&0.05822& 226.5&  10.3&  \nod&0\\
8549&12705&1&  5.7&  5.0& 16.6&241.90721&$+$45.06537&0.04423& 127.1&  10.0&  \nod&0\\
    &     & &     &     &     &241.90495&$+$45.06548&0.04417& 171.0&   9.9&  39.4&3\\
8554& 3702&1&  2.4&  3.7&254.1&182.66434&$+$36.61448&0.08301& 278.3&  10.4&  \nod&0\\
    &     & &     &     &     &182.66413&$+$36.61385&0.08216& 216.2&  10.1&  \nod&0\\
8566& 9102&1&  5.2&  8.5&486.0&114.37415&$+$41.73723&0.08698& 212.4&  10.0&  \nod&0\\
    &     & &     &     &     &114.37248&$+$41.73798&0.08860& 215.4&   9.9&  39.1&1\\
8566&12701&2&  8.5& 15.3& 31.2&114.88098&$+$39.78890&0.09766& 227.6&  10.1&  \nod&0\\
    &     & &     &     &     &114.88025&$+$39.78661&0.09777& 169.7&   9.7&  \nod&0\\
8567&12705&1& 13.6& 15.5& 95.2&120.66084&$+$48.53919&0.05858& 202.8&  10.2&  \nod&0\\
    &     & &     &     &     &120.65726&$+$48.54214&0.05890& 170.8&   9.9&  \nod&0\\
8591& 9101&1&  6.0&  8.6&151.6&212.25921&$+$54.55619&0.07625& 171.5&  10.0&  \nod&0\\
    &     & &     &     &     &212.25641&$+$54.55585&0.07574& 114.5&   9.5&  38.4&1\\
8597& 9101&1&  9.4& 11.7&121.4&224.29840&$+$49.62901&0.06507& 253.1&  10.4&  \nod&0\\
    &     & &     &     &     &224.30071&$+$49.62688&0.06466&  98.7&   9.8&  \nod&0\\
8600& 9101&1&  6.5& 15.8&274.6&244.15762&$+$42.44880&0.13676& 291.7&  10.1&  \nod&0\\
    &     & &     &     &     &244.15996&$+$42.44825&0.13767& 206.5&   9.7&  \nod&0\\
8601&12701&2&  8.9& 15.6&108.7&247.71802&$+$41.28615&0.09395& 188.0&   9.9&  \nod&0\\
    &     & &     &     &     &247.72130&$+$41.28634&0.09359& 132.3&   9.5&  40.1&4\\
8601&12704&1&  4.0&  7.5&164.3&249.07906&$+$40.42270&0.10277& 276.5&  10.3&  \nod&0\\
    &     & &     &     &     &249.07906&$+$40.42159&0.10332& 194.1&   9.9&  \nod&0\\
8606&12703&3& 13.5& 28.6&176.9&253.88789&$+$38.06483&0.11620& 238.3&  10.2&  \nod&0\\
    &     & &     &     &     &253.88723&$+$38.06856&0.11679& 144.5&   9.3&  \nod&0\\
8613& 9102&1&  5.3&  8.4&226.1&254.30980&$+$34.02405&0.08461& 215.6&  10.1&  \nod&0\\
    &     & &     &     &     &254.30958&$+$34.02550&0.08536& 202.0&  10.2&  \nod&0\\
8613&12704&2&  8.5& 16.6&  5.4&255.44264&$+$35.04482&0.10652& 244.9&  10.0&  \nod&0\\
    &     & &     &     &     &255.44271&$+$35.04718&0.10654& 109.4&   9.3&  \nod&0\\
8623& 3702&3&  5.3&  2.9& 78.2&310.37275& $+$0.13226&0.02696&  79.7&   9.5&  \nod&0\\
    &     & &     &     &     &310.37395& $+$0.13311&0.02670&  71.6&   8.9&  \nod&0\\
8711& 3702&1&  3.7&  4.8&288.7&117.33785&$+$52.35114&0.06625& 271.6&  10.3&  \nod&0\\
    &     & &     &     &     &117.33622&$+$52.35083&0.06721& 276.6&  10.5&  \nod&0\\
8714&12701&1&  9.4& 22.0&222.5&117.95382&$+$46.19286&0.13019& 203.1&  10.1&  \nod&0\\
    &     & &     &     &     &117.95731&$+$46.19387&0.12945&  90.5&   9.1&  \nod&0\\
8714&12702&1&  3.4&  1.5&  2.8&118.68964&$+$46.22028&0.02222&  49.1&   8.9&  38.5&1\\
    &     & &     &     &     &118.68828&$+$46.22011&0.02221&  67.9&   8.6&  38.3&1\\
8716&12705&2&  8.8&  7.0&207.4&123.15015&$+$52.73420&0.04026&  41.3&   8.4&  38.5&1\\
    &     & &     &     &     &123.15293&$+$52.73596&0.04096& 515.9&   7.8&  38.8&1\\
8717& 6103&2&  3.4&  6.9&519.6&118.29130&$+$35.93253&0.11333& 246.6&  10.1&  \nod&0\\
    &     & &     &     &     &118.29017&$+$35.93267&0.11160& 248.9&   9.9&  \nod&0\\
8719&12704&3& 11.6&  5.3& 29.8&121.67879&$+$47.31214&0.02258&  65.6&   8.1&  38.8&1\\
    &     & &     &     &     &121.67482&$+$47.31392&0.02248& 129.6&   7.7&  38.6&1\\
8720&12701&2& 10.5& 22.1&150.3&122.29896&$+$49.01703&0.11659& 186.4&   9.9&  \nod&0\\
    &     & &     &     &     &122.29538&$+$49.01533&0.11710& 138.5&   9.4&  38.6&2\\
8720&12703&2&  4.8&  8.6& 84.8&121.61156&$+$50.11970&0.09524& 232.8&  10.2&  \nod&0\\
    &     & &     &     &     &121.61041&$+$50.12083&0.09496& 199.6&   9.8&  \nod&0\\
8720&12705&2& 11.0& 26.3&518.3&122.07704&$+$49.63348&0.13464& 212.7&   9.8&  \nod&0\\
    &     & &     &     &     &122.07946&$+$49.63610&0.13636& 159.5&   9.5&  \nod&0\\
8721& 3702&2&  4.5&  4.0&  0.6&133.79792&$+$54.92827&0.04523& 139.1&  10.0&  \nod&0\\
    &     & &     &     &     &133.80004&$+$54.92855&0.04523&  64.7&   9.0&  38.7&2\\
8721&12703&3&  5.6& 13.9&387.7&135.42069&$+$56.10501&0.14101& 267.4&  10.1&  \nod&0\\
    &     & &     &     &     &135.42226&$+$56.10629&0.14230& 166.7&   9.2&  39.1&2\\
8725& 6101&2&  4.9& 11.1&149.3&127.25650&$+$45.01676&0.12709& 254.3&  10.2&  \nod&0\\
    &     & &     &     &     &127.25658&$+$45.01541&0.12660&  93.9&   9.3&  \nod&0\\
8726&12704&2& 11.2& 10.2&182.1&116.87970&$+$22.60760&0.04587&  61.9&   8.9&  38.7&1\\
    &     & &     &     &     &116.87657&$+$22.60871&0.04647&  91.3&   9.6&  \nod&0\\
8939& 6104&2&  5.3&  8.8&143.0&127.10574&$+$24.62305&0.08844& 292.2&  10.4&  \nod&0\\
    &     & &     &     &     &127.10413&$+$24.62293&0.08892& 255.6&   9.5&  \nod&0\\
8939& 9101&2&  6.3& 11.2&155.0&125.22771&$+$23.72920&0.09665& 245.4&   9.9&  39.3&3\\
    &     & &     &     &     &125.22589&$+$23.72869&0.09717& 383.3&   9.0&  \nod&0\\
8939&12702&3& 10.6& 18.3&209.2&125.39502&$+$23.20666&0.09386& 228.1&  10.1&  \nod&0\\
    &     & &     &     &     &125.39221&$+$23.20803&0.09317& 151.0&   9.7&  38.6&2\\
8943&12704&2&  7.2&  7.5& 79.6&156.43280&$+$36.02361&0.05421& 141.4&   9.4&  \nod&0\\
    &     & &     &     &     &156.43513&$+$36.02294&0.05394& 414.2&   8.8&  40.7&1\\
8945&12703&2& 10.3& 11.7&478.4&173.98187&$+$48.02146&0.05864& 145.0&   9.9&  39.5&4\\
    &     & &     &     &     &173.98301&$+$48.01871&0.05705& 123.9&   9.6&  \nod&0\\
8946&12705&3&  7.5& 17.3&216.2&170.46171&$+$48.07929&0.12896& 195.6&   9.9&  \nod&0\\
    &     & &     &     &     &170.46203&$+$48.08136&0.12968& 125.8&   9.3&  \nod&0\\
8949& 6101&1&  5.9&  2.6&103.7&195.01769&$+$28.15511&0.02198&  52.8&   8.9&  \nod&0\\
    &     & &     &     &     &195.01761&$+$28.15674&0.02232& 243.9&   8.3&  \nod&0\\
8952&12702&3& 10.2& 12.6&274.5&205.32115&$+$26.27205&0.06399& 235.4&  10.3&  \nod&0\\
    &     & &     &     &     &205.32431&$+$26.27223&0.06307&  97.6&   9.4&  \nod&0\\
8987& 9102&2&  5.9&  5.5& 84.3&137.98349&$+$27.89929&0.04740& 117.9&   9.2&  39.5&1\\
    &     & &     &     &     &137.98176&$+$27.89986&0.04768& 115.4&   8.7&  38.0&2\\
9041&12705&3& 11.7& 17.9&103.3&237.45046&$+$29.17128&0.08135& 156.8&   9.7&  \nod&0\\
    &     & &     &     &     &237.44928&$+$29.16819&0.08169& 177.0&   9.9&  \nod&0\\
9042&12705&2&  5.4& 12.9&581.2&235.02388&$+$28.08044&0.13546& 211.6&   9.9&  \nod&0\\
    &     & &     &     &     &235.02256&$+$28.07951&0.13740& 209.1&   9.9&  \nod&0\\
9049&12701&3& 10.0& 12.5&183.1&246.61963&$+$24.02573&0.06485& 157.8&  10.0&  \nod&0\\
    &     & &     &     &     &246.61694&$+$24.02703&0.06546&  81.9&   9.1&  39.0&2
\enddata
\tablecomments{
Column 1: MaNGA Plate number.
Column 2: MaNGA IFU ID.
Column 3: subsample flag: 1 - Primary Sample, 2 - Secondary Sample, 3 - Color Enhanced Sample.
Columns 4-5: projected separation in arcsec and kpc, respectively.
Column 6: radial velocity offset in \kms.
Column 7-8: right ascension and declination in degrees.
Column 9-12: redshift, stellar velocity dispersion, stellar mass, and [O\,{\sc iii}] luminosity from the nuclear spectrum extracted with a 2.6~kpc-diameter aperture, respectively.
Column 13: emission-line classification code: 0 - retired galaxy, 1 - star-forming galaxy, 2 - starburst-AGN composite, 3 - LINER, 4 - type-2 Seyfert, 5 - type-1 Seyfert.  
}
\end{deluxetable*}

\end{document}